%% file: main.tex
\documentclass[5p,number]{elsarticle}

\usepackage[T1]{fontenc}
\usepackage[utf8]{inputenc}
\usepackage{amsmath}
\usepackage{amssymb}
\usepackage{balance}
\usepackage{listingsutf8}
\usepackage[dvipsnames]{xcolor}
\usepackage{hyperref}
\usepackage{cleveref}
\usepackage{comment}
\usepackage{microtype} 

\usepackage{pgfplots}\pgfplotsset{compat=1.9}
\usepackage[inline]{enumitem}
\setlist[enumerate]{label=(\alph*)}

\usepackage{amsthm}

\Crefname{property}{Property}{Properties}

\newcommand{\req}[1]{\textbf{#1}}

\newcommand{\eflint}[0]{eFLINT}

\begin{document}

\input{assets/ast_commands}
\input{clingo_lstset}
\input{eflint_listings}
\lstset{
caption={},
basicstyle=\ttfamily\lst@ifdisplaystyle\scriptsize\fi
}

\title{Reflections on the design, applications and implementations of the normative specification language eFLINT}

\author[1]{L. Thomas van Binsbergen}
\ead{ltvanbinsbergen@acm.org}
\author[1]{Christopher A. Esterhuyse}
\ead{c.a.esterhuyse@uva.nl}
\author[1]{Tim M\"uller}
\ead{t.muller@uva.nl}
\cortext[cor1]{Corresponding author}
\affiliation[1]{organization={Informatics Institute, University of Amsterdam},
  addressline={Science Park 900},
  postcode={1098 XH},
  city={Amsterdam},
  country={The Netherlands}}


\begin{abstract}
\input{abstract}
\end{abstract}

\maketitle

\input{sections/introduction}
\input{sections/background}

\input{sections/running-example}
\input{sections/definition}
\input{sections/design}

\input{sections/implementation}

\input{sections/discussion}
\input{sections/related_work}
\input{sections/conclusion}
\subsubsection*{Acknowledgments}
Special thanks go to all the students, researchers, and domain experts that have provided feedback over the years.
Additional thanks to the anonymous reviewers for their useful suggestions to improving this paper.

This work has been executed as part of the SSPDDP project supported by NWO (628.009.014), the AMdEX Fieldlab project supported by Kansen Voor West EFRO (KVW00309) and the province of Noord-Holland, and the DMI ecosystem supported by the National Growth Fund.
\bibliographystyle{splncs04}
\bibliography{biblio}
\end{document}

%% file: assets/ast_commands.tex
\newcommand{\true}{\mathbf{tt}}
\newcommand{\false}{\mathbf{ff}}

\newcommand{\te}{\mathit{v}}
\newcommand{\ot}{\mathit{ot}}
\newcommand{\fr}{\mathit{dc}}
\newcommand{\ffr}{\mathit{fdc}}
\newcommand{\afr}{\mathit{adc}}
\newcommand{\efr}{\mathit{edc}}
\newcommand{\dfr}{\mathit{ddc}}
\newcommand{\vdc}{\mathit{vdc}}

\newcommand{\mkset}[1]{\mathbf{#1}}
\newcommand{\vars}{\mkset{vars}}
\newcommand{\refs}{\mkset{refs}}
\newcommand{\domids}{\mkset{type\text{-}ids}}
\newcommand{\instexprs}{\mkset{instance\text{-}exprs}}
\newcommand{\boolexprs}{\mkset{boolean\text{-}exprs}}
\newcommand{\exprs}{\mkset{exprs}}
\newcommand{\elems}{\mkset{elems}}
\newcommand{\taggedelems}{\mkset{instances}}
\newcommand{\values}{\mkset{values}}
\newcommand{\constraints}{\mkset{constraints}}
\newcommand{\domains}{\mkset{domains}}
\newcommand{\types}{\mkset{types}}
\newcommand{\basic}{\mkset{basic}}
\newcommand{\derived}{\mkset{derived}}
\newcommand{\strings}{\mkset{strings}}
\newcommand{\modifiers}{\mkset{modifiers}}
\newcommand{\decls}{\mkset{decls}}
\newcommand{\fdecls}{\mkset{fact\text{-}decls}}
\newcommand{\ddecls}{\mkset{duty\text{-}decls}}
\newcommand{\adecls}{\mkset{act\text{-}decls}}
\newcommand{\edecls}{\mkset{event\text{-}decls}}
\newcommand{\pconditions}{\mkset{post\text{-}conds}}
\newcommand{\specs}{\mkset{specs}}
\newcommand{\envs}{\mkset{envs}}
\newcommand{\states}{\mkset{configs}}
\newcommand{\all}{\mkset{all}}
\newcommand{\mutations}{\mkset{mutations}}
\newcommand{\stmt}{\mkfunction{sc}}
\newcommand{\stmts}{\mkset{stmts}}
\newcommand{\actions}{\mkset{actions}}
\newcommand{\abstractions}{\mkset{abss}}
\newcommand{\queries}{\mkset{queries}}
\newcommand{\scripts}{\mkset{scripts}}
\newcommand{\script}{\mkfunction{script}}
\newcommand{\truths}{\mkset{truths}}

\newcommand{\mkfunction}[1]{\mathit{#1}}
\newcommand{\proj}{\mkfunction{proj}}
\newcommand{\dom}{\mkfunction{dom}}
\newcommand{\deriv}{\mkfunction{deriv}}
\newcommand{\precond}{\mkfunction{pre}}
\newcommand{\postcondc}{\mkfunction{post\text{-}cond}^+}
\newcommand{\postcondt}{\mkfunction{post\text{-}cond}^-}
\newcommand{\postcond}{\mkfunction{post}}
\newcommand{\rebase}{\mkfunction{rebase}}
\newcommand{\myapp}{\mkfunction{app}}
\newcommand{\append}{\mkfunction{append}}
\newcommand{\unit}{\theta^0}
\newcommand{\dod}{\mkfunction{dod}}
\newcommand{\saturated}{\mkfunction{saturated}}
\newcommand{\decomp}{\mkfunction{decomp}}
\newcommand{\enables}{\mkfunction{enables}}
\newcommand{\effectsc}{\mkfunction{effects}^+}
\newcommand{\effectst}{\mkfunction{effects}^-}
\newcommand{\bindings}{\mkfunction{binds}}
\newcommand{\myforeach}{\mkfunction{foreach}}
\newcommand{\exts}{\mkfunction{extends}}
\newcommand{\insts}{\mkfunction{subs}}
\newcommand{\var}{\bf {var}}
\newcommand{\function}{\bf {function}}
\newcommand{\restrictions}{\mkset{restrs}}

\newcommand{\stringset}{\mkfunction{string\text{-}set}}
\newcommand{\tstrings}{\mkset{strings}}
\newcommand{\tspec}{\mkfunction{specs}}
\newcommand{\tdecls}{\mkfunction{decls}}
\newcommand{\intset}{\mkfunction{int\text{-}set}}
\newcommand{\myreff}{\mkfunction{ref}}
\newcommand{\consapp}{\mkfunction{construct}}
\newcommand{\subs}{\mkfunction{assign}}
\newcommand{\tuple}{\mkfunction{tuple}}
\newcommand{\tagged}{\mkfunction{tagged}}
\newcommand{\tproduct}{\mkfunction{tuples}}
\newcommand{\tsum}{\mkfunction{sum}}
\newcommand{\ttaggedelems}{\mkfunction{tagged\text{-}elems}}
\newcommand{\csome}{\mkfunction{some}}
\newcommand{\none}{\mkfunction{none}}
\newcommand{\fdecl}{\mkfunction{fdecl}}
\newcommand{\fddecl}{\mkfunction{fderiv}}
\newcommand{\adecl}{\mkfunction{adecl}}
\newcommand{\edecl}{\mkfunction{edecl}}
\newcommand{\addecl}{\mkfunction{aderiv}}
\newcommand{\ddecl}{\mkfunction{ddecl}}
\newcommand{\dddecl}{\mkfunction{dderiv}}
\newcommand{\spec}{\mkfunction{spec}}
\newcommand{\mytag}{\mkfunction{tag}}
\newcommand{\untag}{\mkfunction{untag}}
\newcommand{\action}{\mkfunction{action}}
\newcommand{\remevent}{\mkfunction{event}^-}
\newcommand{\addevent}{\mkfunction{event}^+}
\newcommand{\create}{\mkfunction{create}}
\newcommand{\terminate}{\mkfunction{terminate}}
\newcommand{\obfuscate}{\mkfunction{obfuscate}}
\newcommand{\cons}{\mkfunction{cons}}
\newcommand{\myexists}{\mkfunction{exists}}
\newcommand{\abs}{\mkfunction{abs}}
\newcommand{\bquery}{\mkfunction{bool\text{-}query}}
\newcommand{\iquery}{\mkfunction{inst\text{-}query}}

\newcommand{\cby}{\mkfunction{cond\text{-}by}}
\newcommand{\dfrom}{\mkfunction{deriv\text{-}from}}
\newcommand{\vwhen}{\mkfunction{viol\text{-}when}}
\newcommand{\swith}{\mkfunction{syncs\text{-}with}}
\newcommand{\fcl}{\mkfunction{fcl}}
\newcommand{\ecl}{\mkfunction{ecl}}
\newcommand{\dcl}{\mkfunction{dcl}}
\newcommand{\fcls}{\mkset{fact\text{-}clss}}
\newcommand{\ecls}{\mkset{event\text{-}clss}}
\newcommand{\dcls}{\mkset{duty\text{-}clss}}
\newcommand{\domcl}{\mkfunction{domcl}}
\newcommand{\domcls}{\mkset{dom\text{-}clss}}
\newcommand{\fex}{\mkfunction{fex}}
\newcommand{\dex}{\mkfunction{dex}}
\newcommand{\aex}{\mkfunction{aex}}
\newcommand{\eex}{\mkfunction{eex}}
\newcommand{\fexts}{\mkset{fact\text{-}exts}}
\newcommand{\dexts}{\mkset{duty\text{-}exts}}
\newcommand{\aexts}{\mkset{act\text{-}exts}}
\newcommand{\eexts}{\mkset{event\text{-}exts}}
\newcommand{\fext}{\mkfunction{ext\text{-}fact}}
\newcommand{\dext}{\mkfunction{ext\text{-}duty}}
\newcommand{\aext}{\mkfunction{ext\text{-}act}}
\newcommand{\eext}{\mkfunction{ext\text{-}event}}
\newcommand{\ext}{\mkfunction{ext}}
\newcommand{\extensions}{\mkset{extensions}}
\newcommand{\dblock}{\mkfunction{decl\text{-}block}}
\newcommand{\phrases}{\mkset{phrases}}
\newcommand{\phrbl}{\mkfunction{phrase\text{-}block}}
\newcommand{\phrseq}{\mkfunction{seq}}
\newcommand{\trigger}{\mkfunction{trigger}}
\newcommand{\myforall}{\mkfunction{forall}}
\newcommand{\open}{\mkfunction{open}}
\newcommand{\close}{\mkfunction{close}}

\newcommand{\myeq}{\mkfunction{eq}}
\newcommand{\mygeq}{\mkfunction{geq}}
\newcommand{\where}{\mkfunction{when}}
\newcommand{\mycount}{\mkfunction{count}}
\newcommand{\present}{\mkfunction{holds}}
\newcommand{\enabled}{\mkfunction{enabled}}
\newcommand{\violated}{\mkfunction{violated}}
\newcommand{\opsome}{\mkfunction{some}}
\newcommand{\mysum}{\mkfunction{sum}}
\newcommand{\mynot}{\mkfunction{not}}
\newcommand{\myand}{\mkfunction{and}}
\newcommand{\myor}{\mkfunction{or}}
\newcommand{\project}{\mkfunction{project}}

\newcommand{\mapdelete}{\setminus\!\!\!\setminus}

%% file: clingo_lstset.tex
\lstdefinelanguage{clingo}{
	alsoletter={:-.=1234567890\#<+;},
	sensitive=true,
    breaklines=true,
	stringstyle=\textcolor{brown},
	morestring=[b]",
	morecomment=[l][\color{ForestGreen}]{\%},
	morekeywords=[1]{:-,:,;},
	keywordstyle=[1]\bfseries\textcolor{blue},
	morekeywords=[2]{Person,Agent,Controller,Dataset,Day,A,B,C,D,E,F,I,G,M,N,P,Q,S,U,X,Y,Actor,Bidder,Obj,Price},
    otherkeywords={},
	keywordstyle=[2]\bfseries\textcolor{ForestGreen},
	morekeywords=[3]{0,1,2,3,4,5,6,7,8,9,10},
	keywordstyle=[3]\bfseries\textcolor{brown},
	morekeywords=[4]{=,not,<,+,-},
	keywordstyle=[4]\bfseries\textcolor{purple},
	morekeywords=[5]{\#true,\#count,\#sum,\#min,\#max,error},
	keywordstyle=[5]\bfseries\textcolor{olive}
}

%% file: eflint_listings.tex
\definecolor{mydarkgreen}{RGB}{0,120,0}
\lstset{ 
  backgroundcolor=\color{white},   
  basicstyle=\scriptsize\ttfamily,        
  breakatwhitespace=true,         
  breaklines=true,                 
  captionpos=b,                    
  deletekeywords={data},            
  extendedchars=true,              
  firstnumber=1,                   
  frame=none,	                   
  keepspaces=true,                 
  keywordstyle=\color{blue},       
  morekeywords={Fact, Var, Bool, Function, Event, Physical, Act, Duty, Identified, Derived, When, Where, Holds, Present, Placeholder, For, Not, Create, Terminate, Obfuscate, Created, Enabled,
     Actor, Recipient, Holder, Claimant, Related, Violated, Conditioned, Creates, Terminates, Obfuscates,True, False, Extend, externally, Syncs, with, Extends, Open, Closed,
     from, when, to, by, Sum, Max, Min, Count, Forall, Exists, Foreach, Domain, Force, Predicate,!,+,-,{?},{!?}},            
  numbers=none,                    
  numbersep=1em,                   
  rulecolor=\color{black},         
  showspaces=false,                
  showstringspaces=false,          
  showtabs=false,                  
  stepnumber=1,                    
  tabsize=2,	                   
      commentstyle=\color{mydarkgreen},   
      morecomment=[l]{//},              
  title=\lstname                   
}

%% file: abstract.tex
Checking the compliance of software against laws, regulations and contracts is increasingly important and costly as the embedding of software into societal practices is becoming more pervasive. Moreover, the digitalised services provided by governmental organisations and companies are governed by an increasing amount of laws and regulations, requiring highly adaptable compliance practices. A potential solution is to automate compliance using software. However, automating compliance is difficult for various reasons. Legal practices involve subjective processes such as interpretation and qualification. New laws and regulations come into effect regularly and laws and regulations, as well as their interpretations, are subjected to constant revision. In addition, computational reasoning with laws requires a cross-disciplinary process involving both legal and software expertise.

This paper reflects on the domain-specific language eFLINT developed to experiment with novel solutions to these challenges. Specifically, the language has been developed to experiment with the abstract syntax and semantics of a language supporting different types of reasoning for various applications. The language combines declarative and procedural elements, formalises connections between legal concepts and computational concepts, and is designed to automate compliance checks before, during and after a software system runs. The various design goals and applications areas for the language give rise to (conflicting) requirements. This paper presents and reflects on the current design of the language by recalling applications and requirements. As such, this paper reports on results and insights of an investigation that can benefit language developers within the field of automated compliance.   

{\small \em This is the arXiv version of the paper published by Elsevier under a modified title at \href{https://doi.org/10.1016/j.cola.2026.101411}{https://doi.org/10.1016/j.cola.2026.101411}.}

%% file: sections/introduction.tex
\section{Introduction}
Laws, regulations, contracts, and other \emph{social policies} serve to regulate social systems by establishing \textit{norms} that determine how actors within the system are expected to behave.
A wide range of automated techniques is available to aid in the enforcement of such social policies within software systems.
In distributed software systems, \emph{system policies} are widespread as means to regulate the behaviour of system components, separating the description of the policy from the implementation of the components to enhance adaptability and transparency.
For example, the XACML access control language is designed to regulate access to resources through access control policies that can easily be reconfigured when relevant social policies change~\cite{xacml3-errata}.
As another example, contracts encode how parties’ rights and obligations evolve over time in response to events and can naturally be formalised as state machines.
Smart contracts implement such state machines to manage the exchange of digital assets~\cite{szabo1997}, akin to the automation of the execution of a contract~\cite{governatori2018} without the need of an intermediary.
However, both examples do not directly embed normative concepts such as obligations and power, limiting their expressivity with respect to social policies in general.
Moreover, there is still a significant conceptual gap between the social policy text and its encoding in a smart contract or access control policy.
Existing languages that close this gap by embedding normative concepts are designed for specific subsets of the law.
For example, Symboleo ~\cite{symboleo2020} is used to formalise and execute contracts whereas Catala~\cite{catala2021} targets computational laws such as tax laws.

A general solution to automating compliance that is tenable from the legal perspective needs to address several requirements.
Firstly, a formal representation of policies is required that is sufficiently general to capture policies from a variety of sources and at different levels of abstraction.
For example, the GDPR privacy regulation is more abstract than an organisational policy or a data sharing agreement.

Secondly, the solution should enable different kinds of reasoning.
For example, conformance checking can be used to determine whether an implemented process is compliant with policies based on event logs~\cite{vanderAalst2011}. 
Model checking~\cite{clarkeModelChecking1999} can be used to gain confidence in the correctness of a policy specification.
Simulation helps determine the effects enforcing specific policies on systems or their constituent parts, including human stakeholders.
Search can be used for policy-aware planning and optimisation.

Thirdly, the solution should embrace the fact that legal processes, such as interpretation and qualification, are inherently subjective, and that compliance depends on (in)actions by users.
These observations render it impossible, from a legal perspective, that a system can prevent all violations by design without adaptation or response mechanisms.
This is due to the fact that:
\begin{enumerate*}
    \item
    the same legal norm may be interpreted differently by different stakeholders,
    \item
    conflicting legal norms may simultaneously apply to a case,
    \item
    norms may be interpreted differently and situations may be qualified differently depending on the specifics of a case, and
    \item
    legal obligations may require actions by users that cannot be technically enforced or are not automatically enforced to preserve the users freedom of choice.
\end{enumerate*}
Therefore, solutions are needed in which violations within the system can occur and can be observed, together with penalty- or compensation-mechanisms to respond to non-compliance.
An extensive overview of the challenges related to compliance is given by Hashmi et al.~\cite{Hashmi2018b}.

This paper reports on intermediate findings of an investigation into a general solution to automating compliance.
Central to the investigation is the design and implementation of eFLINT, a domain-specific language (DSL) for developing executable specifications of norms~\cite{eflint}.
The language supports various types of reasoning and can be applied at different stages in the software development process.
The design is based on powerful software engineering concepts such as modularity and extensibility and combines concepts from the declarative, logical, and imperative programming paradigms.
The language has been applied in various experiments demonstrating: assessing individual cases~\cite{eflint}, bounded model checking~\cite{eflint-asp}, run-time compliance checking~\cite{liu2024}, normative multi-agent systems~\cite{10.1007/978-3-031-20614-6_18}, and generating access control policies from social policies~\cite{binsbergen2021b}.
Over time, the language has evolved to become more flexible and suitable for use within these various application areas.
However, the language is oriented towards software experts and is not suitable for use by legal experts.
Instead, the language is intended as an intermediate language that can provide semantics to a surface language such as FLINT~\cite{doesburg2016,breteler2023}, for which it was first developed.

In this paper, we reflect on the design of the language using requirements extracted from applications of the language.
The reflection is based on 11 software systems that incorporate eFLINT to reason about compliance.
The systems are developed by students, researchers and practitioners and demonstrate the effectiveness of the language at various technological readiness levels.
Each system provides either validation of the language in a lab environment or in a realistic test environment with real stakeholder interactions.
In one case, the system provides an operational service maintained by an industrial partner.
These validation experiments are reported in 8 peer reviewed papers, two theses and one white paper (see \Cref{sec:discussion}).
Through these reflections, this paper is intended to be insightful for researchers interested in legal DSLs, norm specification languages, normative reasoning, and automating compliance.
Specifically, this paper contributes:
\begin{itemize}
    \item A list of requirements identified through a range of applications of the language such as static and dynamic compliance checking, verification and planning.
    \item A definition of the latest version of the eFLINT language together with a reflection on the design decisions in relation to the identified requirements. This paper is the first to define the language's extension with \textit{physical actions} and \textit{open types}.
    \item A runtime performance comparison between two implementations of the language: a bespoke reference interpreter and an interpreter that employs Clingo's answer-set solving. 
\end{itemize}

The development of the language went through various phases during which additional applications were targetted and requirements were identified. 
%
%
Rather than attempting to present these as isolated, well-defined phases, we instead present the current design followed by a reflection on the language's evolution, application areas and the requirements identified along the way.  

The paper is organised as follows.
\Cref{sec:background} motivates and gives a high-level overview of the background concepts and terminology most influential on the design.
\Cref{sec:running_example} introduces the language through a running example.
\Cref{sec:definition} defines the abstract syntax and discusses the semantics of the language as it exists today alongside a brief overview of the evolution of the language.
\Cref{sec:design} analyses specific design decisions and trade-offs that have been made to address requirements identified throughout the language's evolution. 
\Cref{sec:implementation} motivates and compares the two most prevalent implementations and includes a runtime performance analysis. 
\Cref{sec:discussion} discusses specific experiments with the language in various application areas that give evidence to the suitability and wide applicability of the language.
In doing so, \Cref{sec:discussion} reflects on the identified requirements, discusses the main limitations of the current design, and suggests future directions.
Related work is discussed in \Cref{sec:related_work} and \Cref{sec:conclusion} concludes.

%% file: sections/background.tex
\section{Background}
\label{sec:background}

This section introduces the main concepts and terminology underpinning the design of eFLINT at a high-level of abstraction.
The section covers process modelling, logic programming and norm representation.
%
%

\subsection{Process Modelling}
Process modelling languages show variety in the purposes for which they are developed and their visual or notational style. 
For example, Business Process Model and Notation (BPMN) is used to model business process using flow-charts~\cite{Aagesen2015}.
In the context of software engineering, the activity diagrams of the Unified Modeling Language (UML) are used to describe the behaviour of a system or system component~\cite{10.5555/1044919}.
On the other hand, languages oriented towards formal analysis, such as nuXmv~\cite{cavadaNuXmvSymbolicModel2014} and Alloy~\cite{Jackson2006Alloy}, tend to use a textual notation. 
Process modelling languages have in common that their semantics can be expressed using event-based mathematical models involving states and transitions.
Examples are petri nets, automata and transition systems.
The choice of semantic model may depend on whether the modelled system has inherent concurrency or parallelism.
Another factor is whether the system can reach infinitely many states or process infinitely many events.

The semantics of eFLINT is described in terms of transition systems in~\cite{eflint} and~\Cref{sec:definition}. 
This approach gives the flexibility to a model a broad range of systems, including systems with infinite, concurrent and bounded parallelism.
For bounded model checking, assumptions can be placed on norm specifications to ensure the underlying transition system has the required, finite structure~\cite{handbook_model_checking}.
Relevant extensions to the semantics of eFLINT that have not yet been considered include additional types of primitives beyond Booleans and integers, probabilistic reasoning and continuous elements, e.g., to represent clocks.

\subsection{Regulatory Compliance}
\label{sec:background-compliance}
In the context of business process modelling, Groefsema et al. define `verification relations' for checking the adherence of system designs and implementations with respect to different types of requirements~\cite{groefsema2022}.
In that work, \textit{compliance} is defined as a relation between a requirement specification and the design of a system, captured by a process model.
On the other hand, \textit{conformance} is a relation between the events logs produced by an enacted process and a requirement specification or design model.
These relations are used to identify `verification techniques' by also considering the moment in the software development lifecycle the checks are performed.
The verification techniques that check adherence with regulatory requirements are \textit{regulatory compliance checking} and \textit{auditing}.

Auditing is verifying the conformance relation between event logs and a formal specification of the regulatory requirements. 
Auditing is an after-the-fact (hereafter: \textit{ex-post}) technique that is applied to complete event logs, typically from a certain time period.
Process mining enables the construction of a process model from the logs~\cite{process_mining_book} as a first step towards verifying the conformance relation. 

Regulatory compliance checking has a static and dynamic counterpart.
The static variant verifies the compliance relation between the model of a system design and the regulatory requirements.
Such a before-the-fact (hereafter: \textit{ex-ante}) technique serves to prevent non-compliance as much as possible.
However, dynamic and ex-post techniques are needed in practice as checking the adherence with certain regulatory requirements might require runtime information, e.g., about user actions and inactions. 
Moreover, the regulatory requirements might themselves be dynamic in nature.  
Dynamic regulatory compliance checking verifies the conformance relation between event logs and regulatory requirements at runtime.

\Cref{sec:discussion} gives an overview of the experiments we performed to demonstrate the use of eFLINT for these types of verification.
In addition, we have demonstrated \textit{adaptability} of systems through the integration of eFLINT.
%

\subsection{Runtime Integration}
\label{sec:reasoners}
%
%
In runtime verification, a component or subsystem is placed under monitoring~\cite{bartocci_introduction_2018}. 
The observed event traces are compared to a behaviour specification typically comprised of a process model describing the intended/permitted behaviour. 
Such a behaviour specification can also encode regulatory requirements~\cite{ceci_toward_2024}.
Observed violations can be used as input to a self-adapting system's MAPE-K loop~\cite{mapek,BONFANTI2023111605} or as part of norm revision~\cite{DellAnna2019}.

In distributed data processing systems, orchestrators assign tasks to worker nodes in fulfilment of a data processing workflow~\cite{9582292}.
Orchestrators plan the workflow by identifying and distributing tasks in adherence to given constraints, which may include regulatory requirements.

In access control, data requests are intercepted to ensure data is accessible only to actors with explicit authorization. 
The XACML access control architecture~\cite{xacml3-errata} defines technical roles for administering, enforcing and deciding on access control policies.
The primary benefit of the separation is that it gives system administrators the ability to update policies without modifying the system's implementation.

In our experiments, eFLINT is integrated in a system by the implementation of an API for interacting with an underlying eFLINT interpreter.
In what follows, we refer to a component implementing such an API as a \textit{reasoner}.

\subsection{Logic Programming}
Process modelling is used by eFLINT to capture the temporal evolution of the state of a system. 
To represent information within the state, eFLINT relies on concepts familiar from logic programming.
Specifically, the user defines sets and relations with predicates determining for every possible element whether it is present in the set/relation according to the current state.
In this context, the state is also referred to as a \emph{knowledge base}.
This style of knowledge representation is common in logic programming languages such as Prolog~\cite{prolog_birth}, Clingo~\cite{DBLP:conf/iclp/GebserKKOSW16} and Datalog~\cite{DBLP:journals/tkde/CeriGT89}, and in ontology languages such as OWL~\cite{Horrocks_Patel-Schneider_McGuinness_Welty_2007,Krotzsch2012} and RDF~\cite{Klyne2004RDF}.
From this paradigm, eFLINT has also adopted \textit{Horn-clauses}; if-then rules for deriving new predicate \textit{instances} from existing predicate instances.
For example, the `counts as' relation is commonly used for sub-typing in ontological reasoning and is easy to implement using eFLINT's derivation rules.
For example, ``if an object $X$ is a car, then $X$ is also a vehicle'':
\begin{lstlisting}[caption={}]
Fact car      // a predicate recognising cars
Fact vehicle  // a predicate recognising vehicles
    Holds when car(vehicle) // cars `count as' vehicles
\end{lstlisting}

The semantics of logic programming languages differ in how they treat Horn-clauses.
For example, Datalog places a restriction on the usage of variables that makes it possible to compute predicates efficiently. 
Prolog and Clingo are more expressive and require more powerful \textit{solvers} such as Potassco for Clingo~\cite{gebser2011potassco}.
Especially relevant to this paper is Clingo and its stable model semantics~\cite{DBLP:conf/iclp/GelfondL88} that enables the specific form of logical programming referred to as \textit{answer-set programming} (ASP)~\cite{Erdem2016}.
A solver for ASP searches for all possible answers -- sets of predicate instances that can consistently hold simultaneously according to the clauses.

Negation is a well-studied challenge in logic programming~\cite{DBLP:conf/adbt/Clark77}.
The use of \textit{negation as failure} within eFLINT is discussed in detail in \Cref{sec:naf}.

The event calculus~\cite{kowalski1986,sadri1995,russo2002,charalambides2005} has been studied extensively as a way to admit temporal reasoning in logics.
Its discrete timepoints and explicitly {evolving} truth-assignment to predicate instances correspond naturally to a transition system semantics in which timepoints correspond to states.
As predicate instances can evolve from being true to false, the event calculus naturally gives rise to a kind of defeasibility of facts, explored separately within defeasible logics~\cite{nute2003,governatori2004}.
In~\cite{eflint-asp}, the semantics of eFLINT is formalised in terms of the event calculus on top of ASP.
%


\subsection{Norm representation}
\label{sec:background-norms}
An important distinction in normative languages is the ability to separate between what can physically be achieved and what is permitted by the norms, i.e., to separate \textit{ability} and \textit{permissibility}. 
An additional separation is needed between permissibility and the concept of institutional \textit{power} as formalised by Jones and Sergot~\cite{jones_sergot_1996_power}.
In essence, an actor can manifest a power by performing an action that has normative implications, e.g., signing a purchase agreement creates an obligation to pay. 
The absence of a permission does not necessarily imply the absence of a power~\cite{doesburg2019}.
An explicit goal for the design of eFLINT has been to support both the concept of institutional power and the deontic concepts of permission, prohibition and obligation.
In general terms, these concepts correspond to those in the legal framework by Hohfeld~\cite{hohfeld1913,hohfeld1917fundamental}.
Note that in this framework the term \textit{duty} is used to refer to the specific `obligation to act' rather than the more general `obligation'.
In this framework, all duties are associated with a `duty-claim' relationship between a duty holder and a duty claimant who can claim benefits from the fulfilment of the duty.
Similarly, all powers are associated with a `power-liability' relationship between a performer and a recipient who receives the effects of the power when applied.

\subsection{Calculemus-FLINT}

The design of eFLINT was motivated to assess scenarios for compliance with norms specified in the Formal Language for the Interpretation of Normative Texts (FLINT) of Van Doesburg~\cite{doesburg2019,breteler2023}.
The intention is for FLINT to be used as a surface language by legal experts and eFLINT as a back-end language by software experts.
%
%
%
%
%
%

The Calculemus-FLINT framework~\cite{doesburg2016} provides a\linebreak{}methodology for interpreting sources of norms to produce a structured, formal interpretation.
The method suggests to recognise actions, duties, actors and facts based on sentence structure and to fill action-, duty-, and fact-\emph{frames} with text fragments extracted directly from the sentences.
An example frame is shown at the top of~\Cref{fig:act-frame}. 
%
%
%
%
%
\begin{figure}[t]%
\begin{minipage}{.42\textwidth}%
\scriptsize%
\begin{tabular}{|p{.35\textwidth}|p{.64\textwidth}|}
\hline
\em Act frame & <<assisting with the contracting of a valid marriage>>\\
\hline
\em Actor & [ordinary or priest ...]\\
\hline
\em Object & [marriage attempt]\\
\hline
\em Interested Party & [spouses]\\
\hline
\em ... & ... \\
\hline
\em Creating postcondition & [valid marriage]; ...\\
\hline
\em Terminates postcondition & [marriage attempt]\\
\hline
\end{tabular}%
\end{minipage}%
\hspace{.1\textwidth}%
\begin{minipage}{.5\textwidth}%
\scriptsize%
\begin{lstlisting}[caption={},deletekeywords={with}]
Fact [ordinary or priest ...]
Fact [spouses]
Bool [marriage attempt]
Bool [valid marriage]
Act <<assisting with the contracting 
      of a valid marriage>>
  Actor       [ordinary or priest ...]
  Recipient   [spouses]
  Related to  [marriage attempt]
  Creates     [valid marriage]()
  Terminates  [marriage attempt]()
\end{lstlisting}\noindent%
\end{minipage}%
\caption{A FLINT and eFLINT specification of the same action-type (Act). This simplified example is taken from~\cite{doesburg2019}.}%
\label{fig:act-frame}%
\end{figure}

\begin{figure}[tb]
\begin{tikzpicture}[scale=.65]
  \node[draw] at (-2.5,5) (source) {\begin{minipage}{2.2cm}sources\\of norms\end{minipage}};
  \node[draw] at (5.5,5) (gavel) {\begin{minipage}{2.2cm}understanding\\of norms\end{minipage}};
  \node[draw] at (5.5,1.5) (scales) {\begin{minipage}{2.2cm}narrative/\\scenario\end{minipage}};
  \node[draw] at (-2.5,1.5) (actor) {\begin{minipage}{2.2cm}actions,\\events,\\and objects\end{minipage}};
  \draw [draw, loosely dashed, gray] (-5,7)node [anchor=west] {\bf physical reality} 
                              rectangle (1.4 ,-0.2);
  \draw [draw, loosely dashed, gray] (1.6,7)node [anchor=west] {\bf institutional reality} 
                              rectangle (8.4 ,-0.2);
  \draw[thick,->] (source) -- node[anchor=north] {\emph{interpretation}} (gavel) ;
  \draw[thick,->] (scales) -- node[anchor=west] {\emph{assessment}} (gavel) ;
  \draw[thick,->] (actor) -- node[anchor=south] {\emph{qualification}} (scales) ;
\end{tikzpicture}%
  \caption{Schematic overview of the processes of interpretation, qualification and assessment as taken from~\cite{eflint} and based on Searle~\cite{searle1996}.}
  \label{fig_interpretation}
\end{figure}
When complete, the resulting interpretation is used to assess a particular case for compliance against the formalised interpretation.
However, the facts of the case also need to be established, through a process known in the legal domain as \emph{qualification}~\cite{searle1996}.
The relation between the aforementioned processes is given in \Cref{fig_interpretation}.
Note that interpretation and qualification are subjective, require legal expertise and may be the subject of disputes, e.g., in a courtroom. 
However, by formalising both the interpretation result and the details of the case, unambiguous assessment can be realised.
By making FLINT executable -- eFLINT abbreviates ``executable FLINT'' -- assessment has been automated.
An eFLINT specification has type declarations to give meaning to frames (see~\Cref{fig:act-frame}) and includes a sub-language for defining scenarios.

\paragraph{Towards an intermediate language}
A long-term goal for the development of eFLINT has been to experiment with the abstract syntax and semantics of an intermediate language for reasoning with arbitrary normative sources. 
The high-level design decisions described in previous paragraphs are motivated by this goal.
The hypothesis has been that the language can be used to formalise arbitrary normative sources by adopting the most fundamental normative concepts. 
Moreover, the language can support different types of normative reasoning for different types of compliance checking by adopting a transition system semantics and logic programming.
The language is intended as a step towards an intermediate language that can support multiple source languages, such as FLINT, and can translate into multiple targets, such as Clingo, for various applications.  

\Cref{sec:running_example,sec:definition,sec:design} give a thorough description of eFLINT.
Reflections on the design, implementations, applications, and requirements are given in \Cref{sec:design,sec:implementation,sec:discussion}.

%% file: sections/running-example.tex
\section{Running Example}
\label{sec:running_example}

This section introduces the eFLINT language through an example related to auctioning.
The code is introduced incrementally with code fragments building on top of each other in the order they are given.
The reference interpreter for the language is available online~\cite{eflint_haskell}.

\textit{Type-declarations} introduce sets and relations, each with a domain whose instances receive a truth-assignment in a runtime state indicating whether the instance is an element of the corresponding set or relation.
The following type declarations introduce the initially empty sets \lstinline-bidder-, \lstinline-object-, \lstinline-price-, and \lstinline-display- and the relations \lstinline-bid- and \lstinline+min-price-of+.
The set \lstinline-display- is declared with \lstinline-Var-, indicating at most one instance can hold true for this type, i.e. \lstinline-display- acts as a variable to which an instance is assigned (optionally).
Similarly, the relation \lstinline+min-price-of+ is `functional', i.e. it maps \lstinline-object-s to a unique price as a mathematical, partial function.
%
\begin{lstlisting}[caption={}]
Fact     bidder       Identified by String // any string
Fact     object       Identified by String
Fact     price        Identified by Int    // any int
Var      display      Identified by object 
Function min-price-of Identified by object * price
Fact bid Identified by bidder * object * price * int
\end{lstlisting}\noindent%
The following \textit{statements} define the function by asserting certain instances.
{
\begin{lstlisting}[firstline=16,lastline=18,caption={}]
+min-price-of(Watch, 100).
+min-price-of(Clock, 200).
+min-price-of(Painting, 400).
\end{lstlisting}}\noindent%
Derivation rules can be added to type declarations to infer truth assignments from knowledge about (other) facts.
The following \textit{type extensions} demonstrate this:
%
\begin{lstlisting}[caption={}]
Extend Fact object Derived from min-price-of.object
Extend Fact price  Derived from min-price-of.price
                               ,bid.price  
\end{lstlisting}
The \lstinline-Extend- keyword adds clauses to an existing declaration.  
The last line above determines that every price included in a bid is an instance of \lstinline-price-.
The results of multiple derivation rules accumulate through set union.
The second line thus determines that the asserted minimum prices are also instances of \lstinline-price-.
A derivation rule can be written as a Boolean expression using \lstinline-Holds when-.
The Boolean expression is evaluated in a context in which the fields of a type are bound as variables, i.e., \lstinline-bidder- and \lstinline-price- below.
%
\begin{lstlisting}[caption={}]
Var highest-bid Identified by bidder * price Holds when 
 (Exists bid: bid.bidder == bidder && bid.price == price 
           && bid.object == display.object 
           && (Forall bid': bid'.price <= price 
                When bid'.object == display.object))    
\end{lstlisting}
This definition assumes that the fields \lstinline-bidder- and \lstinline-price- and the variable \lstinline-bid- can be enumerated to evaluate the expression for each combination of instances of these types.

Action-types are fact-types with instances -- referred to as actions -- that can be \emph{performed}. 
The so-called \textit{postconditions} determine the effects of performing actions. 
The action-type below has two fields, the implicit field \lstinline-actor- and \lstinline-object-. 
An instance of the type holds true when its actor is recognised as an \lstinline-auctioneer-.
A performed action that does not hold true raises a violation, but still has its effects.
The effect is the creation of the instance \lstinline-display(object)-, as well as the implicit termination of any other elements of \lstinline-display- owing to its status as a \lstinline-Var-.
%
\begin{lstlisting}[caption={}]
Var auctioneer // responsible for displaying objects
Act start-bidding Related to object 
  Holds when auctioneer(actor)
  Creates display(object)
Extend Fact actor Derived from auctioneer     
\end{lstlisting}

To create (terminate) an instance is to make it true (false).
As such, performing an action results in the transition from one state to another. 
This also holds for \textit{events} -- essentially actions without a performing actor.
Whereas actions are `performed', events and actions are collectively said to be \textit{triggered}.
Derivation rules are declarative in that they do not cause state transitions.
Instead, they can be understood to `close' a state by deriving new instances from the instances created and maintained (i.e., true and not terminated) by the previous transition. 

The actor of an action can also be named explicitly:
%
\begin{lstlisting}[caption={}]
Act place-bid Actor bidder Related to object, price 
 Holds when bidder
 Conditioned by display(object), price > 
  Max(Foreach bid: bid.price When bid.object == object)
 Creates bid(int = 
  Count(Foreach bid: bid When bid.object == object)) 
\end{lstlisting}
This action-type declaration uses a form of constructor application with implicit fields to create an instance of \lstinline-bid-.
The explicit field \lstinline-int- distinguishes bids from the same bidder.
The implicit fields \lstinline-bidder-, \lstinline-object-, and \lstinline-price- are copied directly from the fields of the triggered instance of \lstinline+place-bid+, i.e., \lstinline-bidder = bidder-, \lstinline-price = price-, etc. 
The \lstinline-Conditioned by- clause establishes preconditions for instances to be \textit{enabled} and actions to be \textit{permitted}.
For an instance to be enabled, it must hold true by being created or derived, and all its preconditions must hold.
A prohibition is the negation of a permission and performing a prohibited action results in a violation.
In this example, bids are only permitted on displayed objects and must involve a price higher than any previous bid.

A \emph{physical} action, as opposed to the previous \emph{institutional} actions, holds by default and is permitted unless it \textit{synchronises} with an institutional action that is prohibited.
Unlike an institutional action, a physical action can thus be disabled and permitted.
The intuition is that conditions on physical actions constrain ability, not permission.
The physical action-type below intuitively captures the \emph{qualification} of a bidder raising their hand at an auction as placing a higher bid on the item currently on display.
%
\begin{lstlisting}[caption={}]
Physical raise-hand Syncs with place-bid(
 bidder = actor, object = object, price = 
  min-price-of.price + increment *
   Count(Foreach bid: bid When bid.object == object))
 When bidder(actor) && display(object) && (min-price-of.object == object)  
 
Var increment Identified by Int
Extend Fact price  
    Derived from min-price-of.price + increment
                ,bid.price + increment
\end{lstlisting}\noindent
The consequence of synchronising an action A with an action B is that A inherits all the pre- and postconditions of B, effectively triggering B whenever A is triggered.
The `synchronises with' relation is directional: in the above, \lstinline+raise-hand+ is not triggered if \lstinline+place-bid+ were to be triggered.
The relation is transitive and potentially cyclic\footnote{Cycles are easy to detect and resolve. Semantically, the transitively closed set of synchronised actions is triggered synchronously.}.
In the above, the type \lstinline-increment- is used as a parameter through which it is determined how much higher the current bid is compared to the previous.
The additional derivation rules for \lstinline-price- are needed to enable those specific instances of \lstinline+place-bid+ that can be triggered by the next occurrences of \lstinline+raise-hand+.

A duty-type declaration defines a fact-type with mandatory fields for a duty-holder and a duty-claimant to establish a duty-claim relation in the Hohfeldian legal framework.
A duty-type declaration has zero or more additional fields (such as \lstinline-price- below) and zero or more violation conditions.
A duty raises a violation when it is enabled and when one or more of its violation conditions hold.
%
\begin{lstlisting}[caption={}]
Bool undue-payment-delay
Duty payment-duty Holder bidder Claimant auctioneer 
  Related to price
  Violated when undue-payment-delay()
\end{lstlisting}
%
%
%
In the example, the duty to pay is created when the bidding on a particular object is ended by the auctioneer.
That is, an auctioneer has the \textit{power} to create payment duties.
%
\begin{lstlisting}[caption={}]
Act end-bidding Holds when auctioneer(actor)
 Creates payment-duty(bidder=highest-bid.bidder
                     ,price=highest-bid.price)
  Terminates display
            ,bid When bid.object == display.object
\end{lstlisting}
The \lstinline-Terminates- clauses refers to the variables \lstinline-display- and \lstinline-bid-.
As they are not fields of \lstinline+end-bidding+, the variables are implicitly bound by an occurrence of \lstinline-Foreach-.
As a result, all instances of \lstinline-display- are terminated and all instances of \lstinline-bid- are terminated for which hold that the object referred to in the bid is on display.
The condition on the termination of \lstinline-bid- is realised by the application of the infix operator \lstinline-When-. 
The notion of power can thus be constrained with the use of \lstinline-When- within post-conditions.
The termination of the \lstinline-display- and \lstinline-bid- facts are considered to occur simultaneously.
The \lstinline[]{Terminates} expressions are evaluated in the same context; the object is still on display in the second \lstinline-Terminates- clause.
The same is true for \lstinline-Creates- and \lstinline-Terminates- clauses and, when in conflict, the creation of a fact is favoured over its termination.

The fragment below shows a sequence of statements forming a \textit{script}.
A script includes assertions of creation (\lstinline-+-) and termination (\lstinline+-+), action/event triggers (without prefix), and queries (\lstinline-?-). 
At the end of the script, Bob wins the auction being the last to raise a hand. 
Bob receives the duty to pay 140 for the watch because all bids but the first raise the minimum price of a 100 with increments of 10, following the definition of \lstinline+raise-hand+.
The query at the end confirms the existence of the duty.
%
\begin{lstlisting}[caption={}]
+bidder(Alice). +bidder(Bob). +bidder(Chloe). 
+auctioneer(David).
+increment(20).
start-bidding(David, Watch). // action trigger
raise-hand(Alice). raise-hand(Bob). raise-hand(Alice). 
raise-hand(Chloe). raise-hand(Bob). 
end-bidding(David).
?payment-duty(Bob, David, 100+4*increment). // True
\end{lstlisting}%
%
%


%% file: sections/definition.tex
\section{Language Definition}
\label{sec:language}
\label{sec:definition}
This section introduces the syntax and semantics of the latest version of the language as implemented by the reference interpreter.
The abstract syntax is defined formally in \Cref{fig:exprs,fig:decls,fig:phrases}.
A concrete syntax definition can be found in the source code of the reference interpreter.
The file \verb+Parse.hs+ contains an EBNF-like grammar description using the grammar combinators of~\cite{binsbergen2018a}.
The sections \Cref{sec:exprs,sec:decls,sec:top-levels} describes the syntax of expressions, declarations and top-level constructs informally. 
In parallel, the static and dynamic semantics of the language are described at a high-level.
A formal semantics is out of scope for this paper. 
A formal treatment of the semantics of a subset of the language is found in~\cite{eflint-asp}.
\Cref{sec:language-evolution} gives a rough overview of the language's evolution.

\subsection{Expressions}
\label{sec:exprs}
\begin{figure}[tb]
\begin{equation*}%
\begin{array}{rlcl}%
d \in &\domids & = & \ldots \\
x \in &\vars  & \supseteq &\domids \\
i \in &\mathbb{Z}  & = & \ldots \\
b \in &\mathbb{B}  & ::= & \true \; | \; \false \\
s \in &\strings    & = & \ldots \\
e \in &\elems & ::= & s \; | \; i \; | \; \tuple(\te_1,\ldots,\te_n) \\
\te\in&\taggedelems & ::= & (e, d)\\
t\in&\exprs & ::= & s \; | \; i \; |\; b \; | \; x \\
    &       & |   & \consapp_t(d, t^*)\\
    &       & |   & \consapp_m(d, m^*)\\
    &       & |   & \project(t,x)\\
    &       & |   & \myforeach(x, t) \; | \; \where(t_1,t_2) \\
    &       & |   & \myexists(x, t) \; | \; \myforall(x, t)\\
    &       & |   & \present(t_1) \; | \; \enabled(t_1) \\
    &       & |   & \violated(t_1)\\
    &       & |   & \myeq(t_1,t_2) \; | \; \mygeq(t_1,t_2) \\
    &       & |   & \mycount(t) \\
    &       & |   & \ldots \;\;(\text{basic operators omitted})\\
m \in &\modifiers & ::= & \subs(x,t)%
\end{array} %
\end{equation*}%
\caption{The abstract syntax of \eflint{} expressions.}%
\label{fig:exprs}
\end{figure}
\Cref{fig:exprs} defines the abstract syntax of eFLINT expressions. 
Expressions are formed by string, Boolean and integer literals, variable occurrences and the application of various operators. 
Expressions are further distinguished into \emph{instance expressions} and \emph{Boolean expressions} based on the type of value they evaluate to.
Boolean expressions are those evaluating to a Boolean value and are formed by the comparison operators and operators such as $\myexists$, $\myforall$, and $\violated$.
Instance expressions are those evaluating to a string, integer or tuple `tagged' with a type identifier of a user-defined type, i.e., an instance $v$ is an element of the sort $\taggedelems$.
As the components of a tuple are tagged values, tuples can be seen as records with the tags as field names.
Instance expressions are formed by operators such as $\consapp$, $\project$, and $\myforeach$.
Expressions that evaluate to a string or integer literal without a tag are only acceptable as sub-expressions. 

Type checking ensures that the right kind of expressions are used within expressions and declarations. 
Automatic coercions can be applied to correct any type mismatches.
For example, the $\present$ operator is automatically applied to an instance expression in places where a Boolean expression is expected.
Sub-expressions evaluating to a string or integer literal are tagged with the correct type identifier in places where an instance expression is expected.
Such coercions can only be applied in places where the expected type identifier can be inferred from context, e.g., within constructor applications.
Coercions may also remove the tag, e.g., in a sub-expression where an integer is expected.

Type identifiers play multiple roles: besides giving a name to a type, they are also used as constructor names, field names, parameters and variables more generally. 
In the concrete syntax, fields, parameters and variables are type identifiers decorated with an index or with zero, one or more prime symbols.
For example, the variables \lstinline-object-, \lstinline-object2-, \lstinline-object'- refer to values of type \lstinline-object-.
This is formalised by the requirement that there is a projection $\proj{} : \vars\rightarrow\domids$ from variables to type identifiers.

Constructor application appears in two forms.
In both cases, a type identifier $d$ is applied as a constructor to create a tuple tagged with the type $d$. 
The form $\consapp_t$ expects as many expressions as arguments as the type $d$ has fields, given in the order matching the declaration of the type.
The form $\consapp_m$ can receive fewer arguments, including none at all, written in any order.
In this form, arguments explicitly associate an expression with a field of the type $d$ (see $\modifiers$).
Any field $x$ not mentioned within an argument of the constructor application is implicitly defined through the modifier $\subs(x, x)$.
The projection operator $\project(t,x)$ is as usual: it expects an expression $t$ that evaluates to a tuple tagged with a type that has the field $x$. 
The value of that field is the result of the expression.
The $\myforeach$ operator is given a variable $x$ and an instance expression $t$ and generates $n$ bindings for $x$, one for each\footnote{When the type $\proj{}(x)$ has an infinite domain, only the instances of the type that hold true in the current state are enumerated.} possible instance $\te$ of the type $\proj{}(x)$.
The expression $t$ is evaluated $n$ times, each time in a modified context binding $x$ to one of the instances $\te$.
The result is a sequence of $m\leq n$ instances.
Instance expressions that are not guaranteed to return a single instance are only acceptable in certain places such as postconditions (e.g., \lstinline-Create-) and as the immediate sub-expression of an accumulator such as \lstinline-Count-. 
The operator $\where(t_1,t_2)$ is available to filter out results in such expressions.
The operator expects an instance expression $t_1$ and a Boolean expression $t_2$ and returns the value of $t_1$, but only when $t_2$ evaluates to true.
As both sub-expressions are evaluated within the same context,\footnote{The context of expression evaluation consists of the set of active bindings and a set of truth-assignments to instances.} $\where$ serves as a guard to ensure that instances are returned only when certain conditions are satisfied.
The operator can only be used in places where an expression is accepted that may return multiple instances.

The $\myexists$ and $\myforall$ operators are evaluated similarly to $\myforeach$ except that they operate on Booleans.
The evaluation of the sub-expression returns zero, one, or more Boolean values.
The operator returns true if and only if one of these values is true, in the case of $\myexists$, or if all do, in the case of $\myforall$.

The operators $\present$, $\enabled$, and $\violated$ are predicates that determine whether the instance computed from their sub-expression satisfies a certain property.
An instance $\te$ is considered to \textit{hold} if $\te$ is assigned `true' in the input to the program, is true in the current state, or is derived within the current state (in that order of priority). 
An instance $\te$ is considered \textit{enabled} if it holds and if all the preconditions (\lstinline+Conditioned by+ clauses) evaluate to true in the current context.
An instance $\te$ is considered \textit{violated} if it is an enabled instance of a duty-type for which one or more of its violation conditions evaluate to true. 

The reference interpreter supports various comparison operators such as $\myeq$ and $\mygeq$ for equality and `greater or equals'.
In addition to \lstinline-Count-, accumulator operators such as \lstinline-Max- and \lstinline-Sum- are available for integer expressions. 
The immediate sub-expression of an accumulator must always be an occurrence of $\myforeach$.
\Cref{sec:naf} provides a detailed discussion on the semantics of the negation operator.

The language is statically typed. 
The type of all variables and user-defined types are known at their declaration and binding sites. This is in part due to the projection $\mathit{proj}$ described earlier that maps a variable name to a type. 
A standard, static type checking algorithm can thus be used to type-check expressions and specifications.
The language is not strongly typed.
A literal value or tuple (see $\elems$) can be used as an instance of more than one type.
Runtime values are therefore tagged with types (see $\taggedelems$).

\subsection{Declarations}
\label{sec:decls}
The core of an eFLINT specification is a collection of type declarations and type extensions.
The syntax of declarations and extensions are defined in \Cref{fig:decls}.
There are four kinds of types that can be declared: facts, duties, actions and events.
Each kind of declaration gives a name $d$ to the type and associates a domain $\delta$ with the type.
\begin{figure}[tb]
\begin{equation*}
\begin{array}{rlcl}
\delta\in&\domains& ::= &\strings \\
       &      & |   & \stringset(s^*) \\
       &      & |   & \mathbb{Z} \\
       &      & |   & \intset(i^*) \\
       &      & |   & \tproduct(x^*) \\
\fr\in&\decls&::=& \ffr \;|\; \afr \; | \; \efr \; | \; \dfr \; \\
r\in&\restrictions& ::= & \var{} \; | \; \function{}\\
\ffr\in&\fdecls&::=& \fdecl(d,\delta^?,r^?) \\
\dfr\in&\ddecls&::=& \ddecl(d,x_1,x_2,x^*)\\
\afr\in&\adecls&::=& \adecl(d,b,x_1,x^*)\\
\efr\in&\edecls&::=& \edecl(d,x^*)\\[.5em]

\ext\in&\extensions &::=& \fex \;|\; \dex \;|\; \aex \;|\; \eex \\
\fex\in&\fexts&::=& \fext(d, \fcl^*, \domcl^*)\\
\dex\in&\dexts&::=& \dext(d, \dcl^*)\\
\aex\in&\aexts&::=& \aext(d, \ecl^*)\\
\eex\in&\eexts&::=& \eext(d, \ecl^*)\\[.5em]

\domcl\in&\domcls&::=& \intset(i^*) \; | \; \stringset(s^*)\\
\fcl\in&\fcls &::=& \cby(t) \; | \; \dfrom(t)\\ 
       &      & | & \open{}() \; | \; \close{}()\\ 
\dcl\in&\dcls &::=& \fcl \; | \; \vwhen(t) \\ 
\ecl\in&\ecls &::=& \fcl \; | \; c \; | \; \swith(t)\\ 
c\in&\pconditions& ::= & \create(t) \; \\
    &            &  | & \terminate(t) \; \\
    &            &  | & \obfuscate(t)%
\end{array}%
\end{equation*}%
\caption{Abstract syntax of \eflint{} declarations.}%
\label{fig:decls}%
\end{figure}%

In the case of a fact-type, this domain can be a set of string literals, integer literals, or tuples. 
A domain made out of a set of literals can be infinite ($\strings$ or~$\mathbb{Z}$) or fully enumerated by the user ($\stringset$ or $\intset$).
The third component of a fact-type declaration determines whether a restriction \lstinline-Var- or \lstinline-Function- is applied.
For fact-types with the \lstinline-Var- restriction, only one instance can hold true at a time, so other instances are automatically terminated when an instance is created.
For fact-types with the \lstinline-Function- restriction, instances are automatically terminated to ensure the binary relation defined by the fact-type is a partial function.
If a domain $\delta$ is omitted for a fact-type declaration, the domain is implicitly the empty tuple, used for fact-types declared via the \lstinline-Bool- keyword. 
Each such fact-type defines a nullary relation.
Effectively, these are Boolean variables, as their only instance receives a Boolean truth assignment in the runtime state.
Fact-types defined with \lstinline-Fact- and without \lstinline-Identified by- have the domain $\strings$ by default.

Duty-, action-, and event-types are associated with a domain of tuples with the fields $x_1,\ldots,x_n$ given as the last $n$ arguments to $\ddecl$, $\adecl$, and $\edecl$ respectively.
In the case of a duty-type, $n$ is $\geq 2$ and $x_1$ refers to the type identifying the \textit{holder} of the duty and $x_2$ refers to the type identifying the \textit{claimant} of the duty.
A duty-type definition thus directly establishes duty-claim relationships.
Fact-types are used to model actors in a system; the abstract syntax does not distinguish actor-types from fact-types.

In the case of an action-type, $n$ is $\geq 1$ and $x_1$ refers to the type identifying the actors that can perform the action.
Any power-liability relationships introduced by the type are derived indirectly by identifying the actors potentially affected by the postconditions.
Note that in the concrete syntax, the actor field is optional and defaults to \lstinline-actor- if not explicitly mentioned.
For example, in the declaration of \lstinline+start-bidding+ the actor field is implicit, whereas in \lstinline+place-bid+ it is explicated as \lstinline+bidder+.
An optional \lstinline-Recipient- clause is also available but is considered deprecated at present.
The Boolean argument to $\adecl$ determines whether the declaration defines a physical or institutional action-type.
A physical action holds true by default as if the action-type has an implicit clauses \lstinline-Derived from True-. 
For an example, see the definition of \lstinline+raise-hand+ in \Cref{sec:running_example}.
An event-type is a physical action-type without an associated actor.
Physical actions and events do not raise violations but may synchronise with institutional actions that do.
A type defined by an earlier declaration within a specification can be re-defined/overridden by a later declaration within a specification. 

A type extension adds clauses to an existing type.
Which clauses can be added to a type depends on the kind of type.
All types can have preconditions ($\cby$) and derivation rules ($\dfrom$). 
Note that \lstinline-Holds when- is syntactic sugar for \lstinline-Derived from- and does not occur in the abstract syntax. 
All types can have open and closed clauses.
Only duty-types can have violation conditions (e.g., \lstinline+payment-duty+).
Only action- and event-types can have postconditions and synchronisation clauses (e.g., \lstinline+raise-hand+).
Only fact-types can have domain clauses.

All clauses are `accumulating' in the sense that zero, one or more clauses of the same kind can be associated with a type.
Each kind of clauses has a specific resolution strategy to handle the absence or multiplicity of clauses of this kind.
Preconditions are conjunctive; all need to be true for an instance of a type to be enabled.
Derivation rules are disjunctive; all instances of a type computed from derivation rules of that type are considered to hold. 
A type is considered closed unless there are only $\open{}()$ and no $\close{}()$ clauses associated with the type. 
A type is thus closed by default, can be opened once, and can subsequently be closed forever. 
Violation conditions are disjunctive; only one suffices for a duty to be violated.
Domain clauses are used to either (a) add new string or integer literals to the already finite domain of a fact-type or (b) reduce the infinite domain of a fact-type to a finite enumeration. 
For example, the following type extension enumerates possible prices of the previously infinite type \lstinline-price-.
\begin{lstlisting}[caption={}]
Extend Fact price Domain 100, 200, 300, 400, 500
\end{lstlisting}
%
%

Semantically, type declarations and extensions define a transition system.
A state in this transition system is a finite set of truth-assignment (true or false) to instances of the types.
Well-typed instances not mentioned in the state are implicitly `unknown'. 
Tested but unknown assignments default to false for closed types or to a runtime exception for open types (further described in \Cref{sec:open-types}).
As such, the set of possible states of the transition system is determined by the domains of the types.
The postconditions and synchronisation clauses of action- and event-types determine the transitions of the transition system by modifying the truth-assignments of the current state to obtain a next state.
A transition is labelled with the synchronised actions and events that together manifest the effects required to realise the transition between the two states.
A \emph{trace} is a sequence of transitions from some starting state.

The effects of postconditions and synchronisation clauses manifest themselves in parallel: when an action $a$ is triggered, all instances computed from the expressions of the postconditions of $a$ are created/terminated and all actions with which $a$ synchronises are triggered.
In this way, a triggered action can cause a chain-reaction of effects resolved synchronously.
The \textit{obfuscate}(t) postcondition is a special case of \textit{terminates}(t) that removes any truth-assignments to the instances computed from $t$ rather than assigning them false.
This variant is needed when working with open types.
The effects of actions and events include possible violations: a triggered institutional action is considered violating if it was not enabled at the moment it was triggered.

The subset of the language with only type definitions, postconditions and synchronisation clauses can be considered the imperative subset of the language, useful for describing process models.
Orthogonally, the subset of the language with only derivation rules can be considered the logic programming subset of the language, useful for inferring new knowledge from existing knowledge.
Finally, the subset of the language that adds violation conditions to duties and preconditions to institutional actions is the normative part of the language.
These clauses give normative meaning to specifications by labelling states and transitions as violating respectively.

\subsection{Top-Levels and Language Evolution}
\label{sec:top-levels}
\label{sec:language-evolution}
\input{assets/table_evolution}
\begin{figure}[tb]
\begin{equation*}
\begin{array}{rlcl}
q\in&\queries  & ::= & \bquery(t) \\
     &         & |   & \iquery(t) \\
\stmt\in&\stmts & ::= & \trigger(t) \; | \; c\\
\script\in &\scripts  & ::= & (\; \stmt \;| \;q\;)^*\\
\spec\in&\specs &::=& (\;\fr \;|\; \ext\;)^*\\
p\in&\phrases &::=& \stmt \; | \; q \; | \; \phrseq(p_1,p_2) \\
    &         & | & \dblock(\spec) \\
    &         & | & \phrbl(p^*) \\
&\mkset{programs} &=& (\spec, \domcl^*, i^*, \script)\\
\end{array}%
\end{equation*}%
\caption{Abstract syntax of \eflint{} top-level constructs.}%
\label{fig:phrases}%
\label{fig:top-levels}
\end{figure}%

\Cref{tab:evolution} gives an overview of the changes between the most notable versions of eFLINT.
%
%
%
%
%
The referenced top-level constructs are formally defined in \Cref{fig:top-levels}.

\paragraph{eFLINT-v1}
The original and primary goal of eFLINT has been to assess concrete scenarios for compliance.
A scenario is a sequence of statements and queries. 
The set of all scenarios is called $\scripts$ in \Cref{fig:top-levels} for consistency with~\cite{eflint}. 
Statements either trigger actions/events or directly create/terminate/obfuscate instances.
Given a specification, a well-typed sequence of statements identifies a trace in the transition system underlying the specification.
Queries are expressions evaluated in the context of the current state and either produce a Boolean result ($\bquery$) or a collection of instances ($\iquery$).

Rather than a specification and a scenario, eFLINT-v1 accepted \textbf{programs} as defined in \Cref{fig:top-levels} (modulo the changes to other syntactic sorts made since).
In addition, a program admits a sequence of domain-related clauses ($\domcls$) to make finite the domains of some or all types of the specification. 
Moreover, a program contains a sequence of instances to explicitly introduce an initial state, considering all the instances mentioned in the sequence as initially holding true.
This specific separation was motivated by a one-to-many relationship between the different components: the same specification can be refined in multiple ways, with multiple initial states defined and multiple scenarios executed.
%
%
%
%
The interpreter for eFLINT-v1 could also be used for simulation by loading the interpreter with a program and stepping through a scenario manually. 

\paragraph{eFLINT-v2 and -v3}
Where programs are useful to assess the compliance of hypothetical and historical scenarios, something more flexible was needed to assess the compliance of dynamically evolving scenarios at runtime.
Programs are especially limited in situations where the specification of the scenario is evolving.
Following the theory of~\cite{binsbergen2020a}, the language was redesigned with eFLINT-v2 to enable the incremental development of a specification and scenario through the execution of \textit{phrases} (see \Cref{fig:top-levels}). 
Phrases enable declarations and statements to be mixed arbitrarily. 
This version also introduced the notion of type extensions.
In eFLINT-v1, declarations had a fixed set and amount of clauses per kind of type.
Since eFLINT-v2, a type can accumulate clauses via type extensions. 

The flexibility of the phrases simplified the implementation of reasoner APIs on top of the interpreter.
For example, in~\cite{10.1007/978-3-031-20614-6_18}, an architecture is presented for normative reasoning in actor-oriented programming, enabling actors that modify their interpretation of norms. 
The application of eFLINT reasoners triggered various language and implementation changes in eFLINT-v3 such as instance queries and open types whose values are determined by the environment rather than the specification. 

\paragraph{eFLINT-v4}
Officially introduced with this paper, eFLINT-v4 can be seen as a consolidation effort in which the syntax and semantics where modified for consistency. 
For example, relative to eFLINT-v3, the semantics of \lstinline-Physical- and \lstinline-Open- were updated according to their definition in this paper.
The only breaking change introduced by eFLINT-v4 is the modification to the semantics of \lstinline-Holds when- to ensure that states are finite, resulting in the current implementation that \lstinline-Holds when- is syntactic sugar for \lstinline-Derived from-.
This change is further discussed in \Cref{sec:design}.

In eFLINT-v4, the top-level syntax is a sequence of phrases that can include \textit{blocks} of phrases.
Semantically, phrases are executed in sequence, with each encountered block having its phrases executed in parallel as one step in the underlying trace.
The execution of a block first distils and processes all the declarations in the block, such that new types are in scope, and then executes all the statements and queries in parallel.
%
%

\paragraph{eFLINT-clingo}
Version eFLINT-clingo was developed for two main reasons: resolving the semantics of negation as failure and enabling search through answer-set programming.
As is discussed extensively in~\cite{eflint-asp}, the reference interpreter does not properly handle negation in some specific situations, resulting in conclusions that are inconsistent with the derivation rules.
The implementation of eFLINT-clingo is an interpreter that internally compiles a specification and a scenario to Clingo.
The Clingo solvers is used for analysing the trace identified by the scenario, e.g., to see which instances are enabled in certain states and which violations occurred.
The output of the Clingo solver is translated back to the user of eFLINT. 

Leveraging the abilities of Clingo, it is possible to search for traces that satisfy criteria.
For example, with a few lines of Clingo added to the generated output, it is possible to find all action-compliant traces that result in duty-violations. 
In future work we aim to add additional top-level syntax to eFLINT-clingo for expressing such search tasks and model checking properties directly in eFLINT syntax.
\Cref{sec:implementation-new} provides details on the features of eFLINT-v4 not yet present in eFLINT-clingo.

As is analysed in \Cref{sec:performance}, eFLINT-clingo is the most performant in almost all situations with respect to runtime.
The variant also received industrial uptake; eFLINT-clingo was chosen by industry partner Abykys for the operationalisation and utilisation of eFLINT to govern data sharing transactions.
\Cref{sec:performance-afsprakenstelsel} provides more details and a runtime performance experiment based on an Abykys use case.

The evolution of the language was driven by additional requirements resulting from various applications of the language.
%
%
\Cref{sec:design} reflects on specific design decisions that were made in response to these requirements.
\Cref{sec:discussion} reflects on the applications and discusses the extent to which these requirements are met.

%% file: assets/table_evolution.tex
\newcommand{\vcolwidth}{0.14\linewidth}
\newcommand{\newVtwo}{
\begin{itemize}[left=-1em, topsep=0em, parsep=0em, partopsep=0em]
\item[] Phrases
\item[] Type extensions
\item[] \lstinline-Syncs with-
\end{itemize}
}
\newcommand{\remVtwo}{
\begin{itemize}[left=-1em, topsep=0em, parsep=0em, partopsep=0em]
\item[] \lstinline-Recipient-
\end{itemize}
}
\newcommand{\newVthree}{
\begin{itemize}[left=-1em, topsep=0em, parsep=0em, partopsep=0em]
    \item[] Instance queries
    \item[] \lstinline-Open- and \lstinline-Physical-
    \item[] \lstinline-Conditioned by-
\end{itemize}
}
\newcommand{\remVthree}{%
\begin{itemize}[left=-1em, topsep=0em, parsep=0em, partopsep=0em]%
\item[] Domain constraints
\item[] \lstinline-Force- action effects
\end{itemize}
}
\newcommand{\newVfour}{
\begin{itemize}[left=-1em, topsep=0em, parsep=0em, partopsep=0em]
    \item[] Phrase blocks
\end{itemize}
}
\newcommand{\remVfour}{
\begin{itemize}[left=-1em, topsep=0em, parsep=0em, partopsep=0em]
\item[] Dynamic \lstinline-Holds When-
\end{itemize}
}
\newcommand{\newVfive}{
\begin{itemize}[left=-1em, topsep=0em, parsep=0em, partopsep=0em]
    \item[] Negation as failure
    \item[] Search
\end{itemize}
}
\newcommand{\remVfive}{

\begin{itemize}[left=-1em, topsep=0em, parsep=0em, partopsep=0em]
\item[] Simplified \lstinline-Enabled- 
\item[] \lstinline-Open- and \lstinline-Physical-
\end{itemize}
}
\begin{table*}[bt]
    \caption{An overview of the evolution of eFLINT. The versions v1 to v4 are referring to different versions of the reference interpreter. Version eFLINT-clingo is an alternative implementation translating eFLINT to Clingo. Row `Top-level' is referring to a syntactic sort in \Cref{fig:phrases}.}
    \label{tab:evolution}
    \centering
    \begin{tabular}{|p{0.1\linewidth}|p{0.11\linewidth}|p{0.13\linewidth}|p{0.16\linewidth}|p{0.18\linewidth}|p{0.16\linewidth}|}
        \hline 
         & \textbf{eFLINT-v1} & \textbf{eFLINT-v2} &\textbf{eFLINT-v3} &\textbf{eFLINT-v4} &\textbf{eFLINT-clingo} \\
         \hline 
         Year                   & 2019              & 2021 & 2023 & 2026 & 2025 \\
         \hline
         Citation               & \cite{eflint}     & \cite{binsbergen2021b} & -- & This work& \cite{eflint-asp}  \\ 
         \hline
         Top-level              & $\mkset{programs}$  & $\mkset{phrases}$        & $\mkset{phrases}$& $\mkset{phrases}$ & $\specs\times\scripts$ \\
         \hline
         New           &    & \newVtwo{}             & \newVthree{} & \newVfour{} & \newVfive{} \\
         \hline
         Removed / Omitted / Deprecated   &   & \remVtwo{}             & \remVthree{} & \remVfour{} & \remVfive{} \\
         \hline 
    \end{tabular}
\end{table*}

%% file: sections/design.tex
\section{Design Reflections}
\label{sec:design}
\input{assets/table_full}
This section reflects on the most important design decisions made in order to better meet requirements identified across eFLINT's evolution.
The requirements are described in~\Cref{tab:requirements1,tab:requirements2} and are referred to in the text in \textbf{bold}.

\subsection{Knowledge Representation}
%
%
\paragraph{Types}
The expressiveness of eFLINT's knowledge representation can be demonstrated in connection to relational algebra.
%
%
%
Relations in eFLINT are algebraic products, akin to database tables, predicates in Prolog~\cite{prolog_birth}, or triples in RDF~\cite{Klyne2004RDF}.
A multitude of derivation rules associated with a type naturally captures the concept of union from relational algebra.
Joins can be achieved by applying \lstinline-Exists-, \lstinline-Forall- and \lstinline-Foreach- operators to enumerate relations and using \lstinline-When- to constrain the result.
The combination of accumulator operators and \lstinline-When- enables grouping.
%

In many programming languages users rely on algebraic sums -- the dual of products -- to represent the concept of heterogeneous collections, e.g., a set of mixed strings and integers (\textbf{datatypes}, \Cref{tab:requirements2}).
Like in Alloy~\cite{Jackson2006Alloy}, heterogeneity is expressed by using the same facts in relations of different types.
    For example, \lstinline|citizen-of(person("Amy"),| \lstinline|country("Germany"))| and \lstinline|member-of(person("Amy"), club("Chess",elo(500)))| treat persons as heterogenous collections mixing memberships and citizenships.

Act-types and duty-types behave as (relational) fact-types in how instances are represented, constructed, created and terminated.
This design choice has made it simple to represent the legal concept of power.
To change the normative positions of (other) actors, an actor can be given actions that create or terminate facts, actions or duties that affect that actor (see \req{normative concepts} in \Cref{tab:requirements1}).

\paragraph{Enumerating Domains}
When using eFLINT for runtime enforcement, the norm specification is likely to require types with \textbf{infinite domains} (\Cref{tab:requirements1}).
After all, the instances involved in a scenario may only become apparent as the scenario unfolds. 
A pragmatic design decision has been made to give semantics to the \lstinline-Foreach-, \lstinline-Forall-, and \lstinline-Exists- operators depending on the domains of the types they enumerate.
The instances of a finite type are all enumerated, independent of whether they hold true or not.
Of an infinite type, only instances that hold in the current knowledge base are enumerated.
By design, enumeration will therefore always terminate. 
In the former case, termination is guaranteed because the domain is finite and only finitely many instances exist.
In the latter case, termination is guaranteed because knowledge bases are finite by design.
The difference is subtle, impactful, and may result in surprising, difficult to diagnose specification errors.
After all, knowledge of the domain of the type is required to predict the enumeration, which is tricky because the domain of a type might be changed by a later extension of the specification.

As an example, the value of \lstinline+count-all+ in the following fragment differs depending on whether the first or second definition of \lstinline-numbers- is used.
With the first definition of \lstinline-number-, all numbers 1 to 5 will be counted.
With the second definition, only numbers 1, 3, and 5 will be counted because \lstinline-Int- identifies the infinite domain of all integers and because only 1, 3, and 5 are in the knowledge base.
%
\begin{lstlisting}[caption={}]
Fact number Identified by 1..5. //finite domain
Fact number Identified by Int.  //infinite domain
+number(1). +number(3). +number(5).
Var count-all Identified by Int Derived from 
    Count (Foreach number: number).
Var count     Identified by Int Derived from 
    Count (Foreach number: number When Holds(number)).
\end{lstlisting}
The definition \lstinline-count- has the \lstinline-When- clause to ensure only numbers that hold are counted irrespective of whether \lstinline-number- was defined with an infinite domain.
This construction can be used to ensure that the enumeration of a type does not change if the domain of the type is modified.
%
%
%

Knowledge bases in eFLINT are finite because instances of types are concrete rather than symbolic, i.e., instances are fully instantiated and do not contain placeholders.
The statement \lstinline-+rich(person)- in the following code fragment determines that every \emph{currently} known person is rich rather than every conceivable, earlier known, or later known person.
As a consequence, Chloe is not considered rich at any moment in the execution of:
%
\begin{lstlisting}[caption={}]
Fact person. 
+person(Alice). +person(Bob). 
+rich(person). +person(Chloe).
\end{lstlisting}
The statement \lstinline-+rich(person)- is to be interpreted as the imperative ``for each known person, assert they are rich'', rather than the declarative ``every person is rich''.
(This statement has an implicit \lstinline-Foreach-, explicated as part of name resolution.)
Note that in applications in which the set of known persons does not change, the declarative and imperative interpretation are consistent.
The decision to maintain concrete facts has been made on the assumption that such knowledge bases are easier to understand for legal experts (\req{explainability} and \req{domain-users}, see~\Cref{tab:requirements2}).
The declarative interpretation can be formulated using a derivation rule to ensure every known person is considered rich, in every knowledge base:
\begin{lstlisting}[caption={}]
Fact rich Derived from person
\end{lstlisting}

The semantics of enumeration also affects derivation rules.
Consider the action \lstinline+start-bidding+ of the running example.
In the following code fragment, the three derivation clauses associated with the action are equivalent.
The first can be seen as syntactic sugar for the second and the second is fully explicated by the third.
%
\begin{lstlisting}[caption={}]
Act start-bidding Related to object // implicit actor
    Holds when auctioneer(actor)
    Derived from 
     start-bidding(actor,object) When auctioneer(actor)
    Derived from (Foreach actor, object: 
     start-bidding(actor,object) When auctioneer(actor))
\end{lstlisting}
The query \lstinline+?Holds(start-bidding(David, Vase))+ does not hold as \lstinline-object(Vase)- is not in the knowledge base.
That \lstinline-object- affects the first clause is counter-intuitive as line 2 does not mention \lstinline-object-. 
\label{sec:alternative-semantics}
In an alternative semantics, we considered binding the field names of the type to the fields of a given instance when determining whether that instance holds true according to a \lstinline-Holds when- clause.
In the example, \lstinline-actor- is then bound to \lstinline-David- and \lstinline-object- to \lstinline-Vase-.
In this case, \lstinline-object- is not enumerated as the variable is already bound and the query holds as \lstinline[]{David} is an auctioneer.
The alternative semantics conflicts with the previous design choice that knowledge bases are concrete and finite.
In the alternative, \lstinline+start-bidding(David,object)+ would hold true for all conceivable objects.
The alternative semantics has been considered and implemented, but has been discarded with eFLINT-v4.

Compared to other logic programming languages, the strategy followed by eFLINT for enumerating the values of types with infinite domains is most comparable to the default strategy for grounding rules in Datalog. 
In Datalog, the possible instantiations considered for variables are determined by the set of true facts of a program. 
This is also the case for the unbounded types of eFLINT.
Note, however, that the current set of known facts can dynamically change in eFLINT, e.g., following the execution of actions and events.
In contrast, eFLINT enumerates all known facts for bounded types, not just those that hold.
Clingo applies a grounding phase to programs, grounding each rule by replacing the rule's variables with all possible instantiations of the variables. 
The domain from which these instances are chosen is determined syntactically, e.g., by including the atoms that occur in rule bodies and heads. 
Datalog, Clingo and eFLINT have in common that they reason `bottom-up' by computing a fixed point from known facts. 
Prolog, as a `top-down' language, reasons from a goal by applying rules in reverse and by using a unification algorithm, thereby avoiding the need to enumerate domains. 
The programmer needs to be mindful when writing rules in order to avoid non-termination caused by, for example, left-recursion or (other) problematic cyclic rules.

\paragraph{Restricting Domains}
\label{sec:restricting-domains}
To dynamically restrict the domain of a type, eFLINT allowed writing \lstinline+When <bool-expr>+ after an \lstinline-Identified by- or \lstinline-Related to- clause. 
%
%
The following (abbreviated) example taken from~\cite{eflint} determines that all valid instances of \lstinline+collect-personal-data+ must satisfy the condition \lstinline+subject-of(subject,data)+.
\begin{lstlisting}[caption={}]
Act collect-personal-data 
 Actor controller Recipient subject 
 Related to data, processor, purpose 
   When subject-of(subject, data) 
 ... // other clauses have been omitted
\end{lstlisting}
Combinations of \lstinline-controller-, \lstinline-subject-, \lstinline-data-, \lstinline-processor-, and \lstinline-purpose- for which the constraint does not hold are thus not members of the type \lstinline+collect-personal-data+ and are therefore not enumerated (e.g., by \lstinline-Foreach- or \lstinline-Forall-). 
This feature was considered crucial in eFLINT-v1 which enabled simulating scenarios (\textbf{micro-simulation}, \Cref{tab:requirements1}).
The interpreter can present (also in eFLINT-v4) users with a finite list of triggerable actions and events.
For each, the interpreter shows whether the action or event is permitted in the current state, thereby showing the difference between \textit{ability} and \textit{permission}. 
The feature to restrict domains served to constrain physical ability separate from compliant behaviour. 
Note that the feature essentially gave eFLINT a data-dependent type system, complicating its semantics.
For these reasons, the feature is considered deprecated since eFLINT-v3.
As an alternative, applications can benefit from the \lstinline-Physical- keyword and present only enabled physical actions to users.
The domains of types can also be statically reduced by extending a type with a \lstinline-Domain- clause (\req{reducible-domains}, \Cref{req:reducible-domains}).

The desired separation between \textit{permission} and \textit{power} motivated the design change in eFLINT-v3 that actions always manifest their effects when executed, even when not permitted.
The keyword to \lstinline[morekeywords={Force}]-Force- the effects of actions, was deprecated by this change.
The construct \lstinline-... When Enabled(...)- can be used to ensure effects manifest only when an action is permitted.

\paragraph{Implicit Arguments}
%
%
In a constructor application with implicit arguments, any missing arguments are implicitly the name of the missing field (see~\Cref{sec:exprs}).
%
%
%
%
Implicit arguments simplify the translation from FLINT to eFLINT.
Consider the FLINT act-frame on the top of \Cref{fig:act-frame} and the corresponding eFLINT code on the bottom.
%
%
The eFLINT code can be automatically generated from the FLINT frame directly.
The resulting eFLINT code is internally consistent and executable.
However, there are some issues with the code.
For example, multiple (pairs of) spouses can be represented as instances of the type \lstinline-[spouses]- but only one valid marriage can be represented as \lstinline-[valid marriage]- is a Boolean type.
This problem can be fixed in at least two ways by redefining some of the types (\req{specialisation}, see \Cref{req:specialisation}).
Firstly, \lstinline-[spouses]- can be defined using the \lstinline-Var- keyword, encoding an assumption that reasoning involves at most one pair of spouses.
Secondly, \lstinline-[valid marriage]- (and \lstinline-[marriage attempt]-) can be re-defined and given a field \lstinline-[spouses]- to determine which pair of spouses are married: 
\begin{lstlisting}[caption={}]
Fact [valid marriage] Identified by [spouses]
\end{lstlisting}
The original act-type definition is still valid with either solution, owing to constructor application with implicit arguments.
In the second solution, the \lstinline-Creates- expression is equivalent to \lstinline-[valid marriage]([spouses])-.
When an action is performed and the expression is evaluated, \lstinline-[spouses]- is bound to the recipient of the action.
The support for implicit arguments and quantifiers in eFLINT thus ensures the direct translation from FLINT to eFLINT yields executable code although extensions may be required to obtain the desired semantics specific to the use case.

The decision which of the two solutions to apply is influenced by the design of the software system in which the code is applied.
If the code is applied to reason about individual, historical cases, then the first solution is effective.
If the code is integrated in a system in which multiple cases are managed, then the second solution might be preferred.
An advantage of the design of eFLINT is that both solutions can co-exist: the original definition in Figure~\ref{fig:act-frame} can be extended by two separate files (using the \lstinline-#require- directive), each implementing one of the solutions. 
A third solution is to run a dedicated (process within an) eFLINT reasoner for each individual case, \textbf{instantiating} (\Cref{tab:requirements2}) the preferred extension with the correct spouses.

\subsection{Service Interactions}
For dynamic regulatory compliance checking of software (defined in~\Cref{sec:background-compliance}), some process is required to automatically convert system events into the steps of a scenario (the \textbf{case input} requirement, \Cref{req:case-input}).
Conversely, the effects of the statements in a scenario need to be observable and interpretable by the system, e.g, to respond to violations (\textbf{case output}, \Cref{req:case-output}).
Initially, a strict separation was maintained between physical processes (external to eFLINT) and institutional processes (internal to eFLINT), relying on code external to eFLINT to make the connection.
However, experience demonstrated that it is convenient to use \lstinline-Physical-, \lstinline-Syncs with- and derivation rules for expressing the required \textbf{qualification} (\Cref{tab:requirements1}) within eFLINT. 
Doing so, eFLINT actions and events model both physical and institutional process, and the two are explicitly connected by eFLINT clauses. 
%
%

\subsubsection{Incremental and Exploratory Programming}
The desire to assess dynamically evolving scenarios and interpretations, motivated one of the most impactful design decisions: with eFLINT-v2, the incremental programming style of languages such as Python was adopted.
In this style, a program is a sequence of individually valid program fragments, referred to phrases in~\Cref{sec:top-levels} and~\cite{binsbergen2020a}.
This is a natural fit with \textbf{event-based} applications and enables \req{case input} to be given as phrases to the interpreter, and its immediate feedback to be emitted as \req{case output}.
For example, the ``raise hand'' button-event can be converted to an eFLINT phrase triggering the corresponding action.
When the auctioneer ends the bidding, the resulting duty to pay can be emitted as a notification to the highest bidder.
The incremental style of programming makes specifications \textbf{extensible}, \textbf{modular} and enables \textbf{dynamic updates} and \textbf{specialisation}.

The change was realised by applying the design methodology presented in~\cite{binsbergen2020a} and implemented using the generic interpreter back-end of~\cite{frolich2021}.
Benefitting from this back-end, eFLINT supports exploratory programming, which enables enables revisiting previous runtime states for debugging (see \req{verification}) and \textbf{micro-simulation}.
The execution history contains information to retrace causes and effects, enhancing \textbf{explainability}, \textbf{transparency} (\Cref{tab:requirements1}) and \textbf{auditability} (\Cref{tab:requirements2}).

For security reasons and to separate concerns, it is beneficial to restrict an eFLINT reasoner to only respond to certain inputs and to make other types strictly internal.
This way, for example, we can disallow any duties from being created or terminated directly, but instead only through institutional actions (powers).
Similarly, we can restrict the creation of authorizations to specific actors and thereby prevent actors from granting authorizations to themselves.
For the running example we might like to say that only the \lstinline+raise-hand+, \lstinline+start-bidding+ and \lstinline+end-bidding+ actions can be triggered.
A possibility is to extend the language with a module-system.
At present, we consider that an eFLINT specification implicitly describes a \textbf{service interface} (\Cref{tab:requirements2}) by exposing only physical actions, events, and open types (see below).
An application-specific eFLINT reasoner can impose further restrictions on the available interactions.

\subsubsection{Open Types}
\label{sec:open-types}
The \lstinline-Open- and \lstinline-Closed- modifiers are available to indicate whether the \textbf{closed world} or \req{open world} (\Cref{tab:requirements1}) assumption holds for the declared type.
The logic for open types is three-valued: true, false or unknown. 
An instance is true or false if and only if it has been explicitly assigned a truth value as part of additional input (described below), or using \lstinline-Creates-, \lstinline-Terminates-, \lstinline-+<EXPR>-, \lstinline+-<EXPR>+, or when it is derived as true.
The truth of all other instances is unknown.
%
%
A closed type can be seen as a special case of an open type for which unknown is also interpreted as false. 
The \lstinline-Obfuscates- clause has been added to the language to remove the truth assignment of an instance.
%
%

The open types of a specification serve as parameters of the specification capturing knowledge only available within the environment in which the specification is applied (\req{service interface}).
As a rule of thumb, one can declare those (fact-)types as open whose truth values are not modified by the specification itself. 
%
%
For the running example, we could decide the following: 
\begin{lstlisting}[caption={}]
Open Function min-price-of Identified by object * price
Open Fact     bidder       Identified by String
Open Var      auctioneer   Identified by String
\end{lstlisting}%
Note that the information captured by open types may be subject to change during execution. 
For example, it may be reasonable to assume additional bidders are added, whereas perhaps the minimum price of an object is fixed once set.
A mechanism has been added to eFLINT-v3 to provide `additional input' alongside a phrase to instantiate certain types for the evaluation of this phrase only.
For establishing truth, input has a higher priority than creation/termination which has a higher priority than derivation.
%

The treatment of unknown affects how applications interact with eFLINT reasoners.
The main design decisions are what happens when (a) evaluation requires the determination whether an unknown instance holds true and when (b) evaluation requires the enumeration of the instances of an open type with an infinite domain.
The execution of a phrase is interrupted, and an exception raised, when an unknown truth-assignment to an instance is required to evaluate a Boolean expression.
The execution of a phrase is also interrupted when the instances of an open type need to be enumerated and the domain of the type is infinite.
For open types with an infinite domain, the truth of all instances not recorded in the knowledge base is unknown and it is therefore unknown which instances require enumeration. 
In this case, the open type itself is reported as part of an exception alongside the interruption.

Both types of exceptions can be understood as a request to the execution context to provide additional input.
The first kind of exception can be resolved by the execution context providing a truth assignment to the reported instance.
The second kind of exception can be resolved by the execution context providing the instances within the domain of the reported type that are to be enumerated.
To resolve exceptions and continue, additional input is provided as a set of truth assignments to instances.
%
%
%
%

Two motivations for closing a type previously declared as open have been encountered.
Firstly, certain static analyses may require finite domains and may require the closed world assumption for all types.
%
%
Secondly, a specification with open types may be extended to specialise towards a particular application. 
For example, an open-type such as \lstinline-bidder- may be refined to include knowledge about valid bidders in the specification. 
Within a particular application it may be valid to say that ``all users except the auctioneer are valid bidders'', expressed as follows:
\begin{lstlisting}[caption={}]
Open   Fact user   Identified by String
Closed Fact bidder Identified by String 
         Derived from user When Not(auctioneer(user))
\end{lstlisting}%
The newly introduced type is typically an open type, such as \lstinline-user- in the example above.
With these definitions, \lstinline-bidder- can be considered `internal' and \lstinline-user- `external'.
%


\subsection{Reasoning by Default}
\label{sec:naf}
The type of reasoning applied to determine the truth assignment of instances is known as \textit{reasoning by default}: instances hold if they were created or derived by some clause, and otherwise, by default, they do not hold.
This semantics imposes an asymmetry between truth (holding) and falsity (not holding):
only truths are witnessed, by creations or derivations.
How can we be certain that at any given moment, all truths have been witnessed to definitively evaluate occurrences of \lstinline-Not-, \lstinline-Foreach-, \lstinline-Count-, etc.?

The literature on logic and logic programming extensively explores reasoning by default (also called \textit{negation as failure})~\cite{DBLP:conf/adbt/Clark77}.
Its definition of falsity as the complement of truth is simple and powerful.
But, as with many logic programming languages before, reasoning in \eflint{} is complicated by the existence of syntactically valid specifications which have no single logical interpretation.
For example, which bidders are ready, according to the definition below?
\begin{lstlisting}[caption={}]
Fact ready Identified by bidder
 Holds when Not(ready(bidder)).
\end{lstlisting}

In general, the issue arises from any case where truth is cyclically dependent on its own falsity.
These dependencies can be more subtle than in the example above.
In the example below, the contradiction of defining minimum prices as being lower than themselves is not as obvious.
This issue can be corrected by replacing \lstinline{<=} with~\lstinline{<}, altering its meaning to specify that no (different) assignments of value to the same object may be larger. 
\begin{lstlisting}[caption={}]
Function min-price-of Identified by object * price
  Derived from bid Where 
    Not(Exists min-price-of':
        min-price-of'.object == bid.object
     && min-price-of'.price <= bid.price).
\end{lstlisting}

Specifications without these troublesome cyclic dependencies are called \textit{stratified} because, intuitively, facts can be categorised into ordered strata, such that the facts depend only on facts in preceding strata, necessarily having no cycles.
Several approximations of this ideal notion of `stratification' have been explored~\cite{DBLP:conf/pods/Ross90}, because it is generally costly to accurately recognise and exploit.

The original work on \eflint{} did not consider the subtleties of reasoning by default.
For many years, the interpreter implemented a `greedy' algorithm, which escaped any cycles by overriding falsity with truth as necessary.
This solution sufficed to enable years of experimentation and application.
Many problems were naturally expressed by stratified specifications.
Consequently, the interpreter usually behaved as expected.
Otherwise, issues were usually discovered quickly, and could be worked around by just enough refactoring until the underlying cycles vanished.
But the reasoning algorithm was found to exhibit another, even more subtle undesirable characteristic:
users could not predict how the algorithm breaks the tie between mutually exclusive truths~\cite{eflint-asp}.
For example, nothing in the following specification helps the reader to predict exactly which bidder becomes the only auctioneer by default.
\begin{lstlisting}[caption={}]
Extend Fact auctioneer Derived from bidder
   When Not(Exists auctioneer': auctioneer' != bidder).
\end{lstlisting}
As with cyclic dependencies, minor refactoring usually suffices to break all ties.
However, in complex specifications, it can be very difficult to recognise the problem and to diagnose the tie or cycle that is to blame.

In \cite{eflint-asp}, we adapted the \textit{answer set} semantics (also called the \textit{stable-model} semantics)~\cite{DBLP:conf/iclp/GelfondL88} for \eflint{}.
Intuitively, it enumerates all possible outcomes from all acyclic interleavings of reasoning steps.
Stratified specifications still have unique answers but unstratified specifications can have no answer - in the case of cycles - or many distinct answers - in the case of ties.
Either case recognises the presence of problems in the specification, which users can then address.

In an on-going, separate research effort we are exploring the \textit{well-founded semantics}~\cite{van_gelder_well-founded_1991} as an alternative to the answer-set semantics (e.g., see the related Seaso language in~\cite{esterhuyse_thesis}).
These semantics are closely related, but they differ in a crucial detail:
like \eflint{}'s original `greedy' semantics, the well-founded semantics ensures that every specification has a unique answer, and like the answer-set semantics, the truth assignment of the answer preserve the constraints imposed by derivation rules.
The trick is that the well-founded semantics assigns unknown as a distinct value to instances, and does so to those instances whose truth or falsity would be inconsistent with the constraints.
The well-founded semantics thus marks and isolates troublesome instances for users to diagnose.
In the prior \lstinline-auctioneer- example, the set of bidders remains clear and determined, but whether each bidder is an auctioneer is unspecified, because any Boolean valuation would arbitrarily prefer one instance over another. 
The well-founded semantics and answer-set semantics align in an intuitive way: truth and falsities in the former have the same value in \textit{all} answers of the latter~\cite{van_gelder_well-founded_1991}.
That is, the well-founded semantics can be understood to `summarise' the answer-set semantics.

%% file: assets/table_full.tex
\begin{table*}[bt]%
\caption{A description of legal and reasoning requirements.}%
\label{tab:requirements1}
\label{req:interpretation}
\label{req:qualification}
\label{req:abstraction-level}
\label{req:open-texture-terms}
\label{req:assessment}
\label{req:ex-ante}
\label{req:ex-post}
\label{req:auditability}
\label{req:forward-chaining}
\label{req:backward-chaining}
\label{req:norm-priorities}
\label{req:infinite-domains}
\label{req:finite-domains}
\label{req:reducible-domains}
\label{req:open-world}
\label{req:closed-world}
\label{req:micro-simulation}
\label{req:macro-simulation}
\label{req:search}
\label{req:specialisation}
\centering
\begin{tabular}{|l|p{0.75\textwidth}|}
\hline
\em legal & The language supports ...\\
\hline
\bf interpretation         & the formal interpretation of a wide range of norms, e.g., laws, regulations, and agreements \\
\bf qualification & the formal qualification of real-world events and objects to institutional actions and facts \\
\bf normative concepts     & the most fundamental normative concepts, e.g., permission, obligation, power, etc. \\
\bf abstraction level      & formalising normative sources at their own level of abstraction \\
\bf open-texture terms     & terms whose meaning is left abstract for context-specific specialisation\\
\bf specialisation         & redefining types to provide a specialised definition with a specific context \\
\bf assessment             & assessing the compliance of scenarios, e.g., legal or social cases \\
\bf ex-ante enforcement    & assessing the compliance of actions (processes) before they are performed (executed)\\
\bf ex-post enforcement    & formalising responses to violations such as penalties and compensation \\
\bf auditability           & recording the information required to establish how compliance decisions were made \\
\hline
\em reasoning & The language supports ...\\
\hline
\bf forward chaining     & applying derivation rules and transitions in `forward' direction \\
\bf backward chaining    & applying derivation rules and transitions in `backward' direction\\
\bf norm priorities      & defining strategies to resolve conflicts between norms \\
\bf infinite domains     & reasoning without enumerating all potentially relevant instances a priori \\
\bf finite domains       & reasoning with a finite set of potentially relevant instances  \\
\bf reducible domains    & modifying the set of valid instances of a domain, admitting/simplifying finite reasoning \\
\bf open world           & reasoning with unknown truth-values \\
\bf closed world         & reasoning with the closed world assumption \\
\bf micro-simulation     & playing out hypothetical scenarios to see the effects of actions and their alternatives \\
\bf macro-simulation     & simulating the effects of interpretations on sets of scenarios to anticipate implementation \\
\bf search               & finding scenarios that satisfy certain (un)desirable properties\\
\hline
\end{tabular}
\end{table*}
\begin{table*}[bt]
\caption{A description of requirements related to services and usability.}%
\label{tab:requirements2}
\label{req:case-input}
\label{req:case-output}
\label{req:service-interface}
\label{req:persistence}
\label{req:dynamic-updates}
\label{req:data-migrations}
\label{req:event-based}
\label{req:instantiation}
\label{req:usability}
\label{req:datatypes}
\label{req:source-references}
\label{req:versioning}
\label{req:verification}
\label{req:modularity}
\label{req:extensible}
\label{req:transparency}
\label{req:domain-users}
\label{req:explainability}
\begin{tabular}{|l|p{0.78\textwidth}|}
\hline
\em services & The language supports ...\\
\hline
\bf case input         & receiving information from the external environment as input \\
\bf case output        & providing information to the external environment as output \\
\bf service interface  & specifying which information can be provided as input to a specification \\
\bf persistence        & persistent storage of all runtime state needed for reasoning  \\
\bf dynamic updates    & modifying the specification at runtime \\
\bf data migrations    & modifying instances to recover their validity w.r.t. a modified type \\
\bf event-based        & responding to events, actions and observations about situations in the environment \\
\bf instantiation      & instantiating specifications within specific contexts, e.g., via a parameter mechanism \\
\hline
\em usability & The language supports ...\\
\hline
\bf datatypes & primitive, composite and user-defined types to represent facts about the world\\
\bf domain-users      & both legal experts and software experts in their purposes for using the language\\
\bf source references & connecting code fragments to fragments from the normative source such as a regulation \\
\bf transparency      & communicating about interpretation and compliance decisions to affected stakeholders \\
\bf explainability    & explaining compliance decisions to relevant stakeholders \\
\bf versioning        & maintaining multiple versions of specifications and scenarios \\
\bf verification      & verifying a specification through testing, debugging, and/or property checking \\
\bf modularity        & establishing specifications and scenarios modularly \\
\bf extensible        & extending existing specifications in a modular fashion, e.g., for specialisation and exceptions \\
\hline
\end{tabular}
\end{table*}

%% file: sections/implementation.tex
\section{Implementations}
\label{sec:implementation}
This section discusses several aspects related to the implementation efforts of the eFLINT language, focussing primarily on the reference interpreter versions 1 through 4 in Haskell and eFLINT-clingo interpreter implemented in Rust. 
The motivations behind the two implementations are discussed and performance comparisons are made.

\subsection{The Reference Interpreter}
\label{sec:implementation-old}
The Haskell implementation of eFLINT~\cite{eflint_haskell} serves as a tool to experiment with the semantics of the language and to prototype with alternatives.
The decision was made to implement eFLINT as a definitional interpreter, thus simultaneously defining and implementing the language.
%
The main requirements for the implementation were code readability and that adjustments should be simple to make.
The evaluation functions central to the interpreter are implemented via a principled usage of well-established Haskell monads: the Reader monad for dealing with bindings and scope, the State monad for state transitions, and the Writer monad for output and exceptions.
The resulting definitions are comparable to a big-step style structural operational semantics and, for the well-acquainted, concisely convey a lot of information about the implemented semantics. 
The grammar combinators of~\cite{binsbergen2018a,binsbergen2020} are used to define the concrete syntax of the language.
The result is a syntax definition that resembles abstract syntax quite closely whilst defining a (GLL) parser directly.
These and other implementation choices promoted the requirements, generally doing so at the cost of performance. 
For these initial purposes, the interpreter was sufficiently self-documenting. 
However, as the user-base started to grow, other demands were being placed on the (tool) support for the language.
Firstly, additional forms of documentation were added.
Jupyter notebooks were developed to give a gentle and high-level introduction of the main language features, curated examples were taken from experiments and use cases, and an ever-growing set of unit tests help testing and documenting expected behaviour of specific language features and corner cases.

Secondly, additional front-ends and means to interact with the interpreter were needed, resulting in REPL and API front-ends replacing the initial simulator front-end.
The REPL front-end replaces the simulator in the most direct sense and on top of the usual REPL interactions offers meta-commands for exploring (alternative) scenarios in a step-by-step fashion.
The API front-end has been used to develop several wrappers for accessing the interpreter as a reasoner, e.g., in Python, Java, and Scala for actor-oriented programming within the AKKA framework. 
Especially for users of the API front-end (and the users of the software they build), the runtime performance of the interpreter renders it impractical beyond basic examples.
For other users, the interpreter serves as a reference implementation to verify their own experimental implementation of the language.
For this reason, we decided to maintain the original requirements for the Haskell interpreter and investigate alternative, more performant implementations.

Thirdly, with more users, eFLINT was exposed to additional scrutiny and over time it was shown that the documentation was not always complete.
Specifically, the Jupyter notebooks and curated examples are too high-level for those seeking a detailed understanding of the language.
%
The test suite provides further details but was repeatedly found to be incomplete and had to be extended as new bugs or corner-cases were discovered.
Most problematically, in some cases the interpreter was discovered to exhibit behaviour that was not expected or desired upon closer inspection.
These cases occurred when implementation choices in the interpreter were not made consciously but rather followed naturally from the way the code was organised or other features were implemented.
This prompted the effort to develop a formal semantics of the language.
\subsection{The eFLINT-clingo Interpreter}
\label{sec:clingo-reasoner}
\label{sec:implementation-new}
From the observations of the previous subsection we concluded that a formal definition of the semantics of the language is needed, triggering the work on a translation from eFLINT to the Clingo language reported in~\cite{eflint-asp}. 
Specifically, this translation and resulting implementation addresses the following points at once:
\begin{enumerate}
    \item The \eflint{} semantics is formally specified to fully document the language's main features in all their details. This point is addressed by specifying a translation from eFLINT to (a subset of) Clingo, inheriting from Clingo's semantics in several regards.

    \item The translation addresses the Haskell interpreter's problems with reasoning by default, opting for the stable model semantics of Clingo.
    
    \item The runtime performance of eFLINT code is significantly improved, especially in regards to the evaluation of derivation rules, benefitting from the years of development and optimisation efforts that went into the Clingo solver. 
    Clingo's superior performance has been demonstrated at international competitions~\cite{DBLP:journals/ai/CalimeriGMR16}.

    \item The implementation supports model-checking and searching for scenarios satisfying given criteria (supporting \textbf{verification}, \Cref{tab:requirements2} and \textbf{search}, \Cref{tab:requirements1}). 
\end{enumerate}

The implementation consists of the translation pipeline and calls the Clingo solver internally before converting its output back to eFLINT.
This has enabled the implementation of a reasoner on top of the eFLINT-clingo interpreter, roughly comparable to the reasoners developed on top of the reference interpreter.
However, at present, eFLINT-clingo does not support all reasoning features.

\paragraph{Unimplemented Features}
\label{sec:removed-reasoning}
In \Cref{sec:design} it was discussed that the reference interpreter adopted incremental programming features, supporting its use in cases where user input is interleaved with reasoning and output.
Firstly, it raises exceptions to request additional input when evaluating instances of \lstinline-Open- types.
Secondly, it accepts type extensions and statements as phrases.

At present, the eFLINT-clingo interpreter supports neither of these features.
A specification and scenario are input first, and evaluated second.
Work continues to support these features by re-introducing a layer that maintains a mutable intermediate state.
The hope is to exploit Clingo's more advanced features and configuration interface~\cite{DBLP:journals/tplp/GebserKKS19}, such that the Clingo solver maintains the intermediate state as much as possible.
In the meantime, users can approximate these features by maintaining an specification as mutable state themselves, and interleaving their mutations with calls to the interpreter.

Another semantic difference is that the \lstinline{Enabled} check on an action $a$ does not involve checking the preconditions of actions that synchronise with $a$ via \lstinline{Syncs with}.
This dramatically simplifies the translation to Clingo and its reasoning about each action in each state.
The original semantics is recoverable by (speculatively) performing the action and checking if a violation results.

There is also no distinction yet between \lstinline-Physical- and institutional actions. 
We do expect to be able to leverage this distinction for static regulatory compliance checking using the Clingo solver (as defined~\Cref{sec:background-compliance}).


\subsection{Synthetic Performance Comparisons}
\label{sec:performance}

In this section, we report on a suite of synthetic performance tests to compare the reference interpreter for eFLINT-v4 and the eFLINT-clingo interpreter.
Recall from \Cref{sec:naf} that these two interpreters produce different outputs given some inputs expressing reasoning by default.
For a fair comparison on performance, we consider only cases where the two interpreters produce the same output.
Evidence of the equivalence of the two implementations in these cases is given in~\cite{eflint-asp}.
We initially focus on concise and synthetic inputs crafted to isolate our testing of the performance characteristics of the interpreters to semantic properties of the input.
For example, one such property is the length of the input scenario.
In \Cref{sec:performance-afsprakenstelsel} we analyse the performance of the eFLINT-clingo interpreter on a real-world example.

With the crafted tests, we have identified trends in the responses of both interpreters that reveal their suitability to various kinds of input.
We find that the eFLINT-clingo's relative strength is in handling cases of complex, yet structured reasoning in each scenario step, and it is relatively weak in reasoning about long scenarios with trivial reasoning in each state.
We ultimately find that \eflint{}-clingo generally outperforms \eflint{}-v4 by a significant margin, and the few cases of slowdown are modest.

\paragraph{Experimental Setup}

\input{clingo_perf1}

\Cref{fig:perf_arith,fig:perf_combo,fig:perf_primes,fig:perf_long_time,fig:perf_chain} plot the results of an illustrative selection of a larger set of experiments available in full at~\cite{eflintClingoImpl}.
Per plot, each measurement is the median of 10 runs on a machine with an Intel core i9 9th generation processor with 16 hardware threads, 32GB RAM DDR4 memory, and an NVMe hard drive.
The experiments compare both interpreters' single-threaded performance.
%
Cursory experiments found that Clingo achieves no speedup from multi-threading for inputs producing just one answer.

\paragraph{Discussion of Experiments}

\Cref{fig:perf_chain} evaluates the performance of the interpreters given an input initialised with many instances, but where only one derivation step is possible at a time.
For example, in any scenario, only \lstinline{x(50)} is derivable just after \lstinline{x(51)} is derived.
The interpreters exhibit similar response curves: runtime grows superlinearly with the number of derivations, presumably as a consequence of paying the price of storing and traversing larger knowledge bases.
However, eFLINT-clingo starts faster, and slows down more gradually.
The net result is that, compared to eFLINT-v4, eFLINT-clingo begins with a $2\times$ speedup and grew to $25\times$ for the largest input that we tested.
We have no explanation for $N = 2^{11}$ representing a significant outlier, but we note that it was reliably reproduced by both interpreters.

\Cref{fig:perf_arith} demonstrated a case where eFLINT-clingo maintained approximately $\times2$ speedup across all test scenarios.
We expect that the cost of the division operator predominates, but more testing is required.  

\Cref{fig:perf_combo} shows the results of an experiment expected to emphasize the strengths of Clingo: the interpreter must quantify, combine, and compare many terms.
This specification causes runtime cost to scale more sharply with the independent variable, but the speedup of the new interpreter over the original are similar to that shown in \Cref{fig:perf_chain}:
the simplest scenario induces $2\times$ speedup, and the largest scales up to $17\times$ speedup.

\input{clingo_perf2}

\Cref{fig:perf_long_time} shows a case where the reference interpreter for eFLINT-v4 is expected to excel:
minimal reasoning is spread over a long scenario.
Intuitively, the reference interpreter was optimised around the `chronological' characteristic of \eflint{} scenarios and states: reasoning about each state~$N$ depends only on states $N-1$ and $N$.
The reference interpreter reasons sequentially, wasting no time considering unrelated states.
By comparison, Clingo threatens to reason na\"ively, because the entire \eflint{} specification and scenario are encoded as a monolithic Clingo ruleset, and we cannot assume that Clingo recognises and exploits its subtle chronological property.
Indeed, we find that eFLINT-clingo's speedup peaks with $94\%$ at $N=2^6$, waning to $82\%$ speedup at $N=2^{10}$.
However, clearly, eFLINT-clingo overall still outperforms eFLINT-v4 by a significant margin, and this experiment suggests that its performance advantage will only be lost in scenarios far longer than we expect to encounter in practice.
Nevertheless, we consider this to be an interesting direction of future research, to which we hope to leverage Clingo's \textit{multi-shot solving}~\cite{DBLP:journals/tplp/GebserKKS19}, where Clingo's processing of input is interleaved with reasoning and output.

Finally, \Cref{fig:perf_primes} shows a case where eFLINT-v4 outperformed eFLINT-clingo.
This input is a carefully crafted encoding of the \textit{prime sieving problem}: enumerate all prime numbers to a given maximum~\cite{DBLP:journals/symmetry/BahigHANB22}.
Here, we expect Clingo's heuristics for selecting rules to fail.
Both interpreters use a similar solution: na\"ively enumerate triples $(X,Y)$ and test if $X \times Y = Z$ for each~$Z$.
Performance then depends on the order that the integer pairs are enumerated.
Evidently, eFLINT-v4 gets lucky more often.
But this outcome was sensitive to the encoding of the problem.
For example, chaining \lstinline{addleq} via \lstinline{Derived from} instead of \lstinline{Syncs with} slowed down eFLINT-v4.
More experimentation is required to precisely explain the sensitivity to such particulars of the specification.
We conclude that eFLINT-clingo benefits from specifications with more structured reasoning, where Clingo's heuristics try applicable rules more often.

\subsection{Realistic Performance Tests}
\label{sec:performance-afsprakenstelsel}

This section is intended to demonstrate that \eflint{} is sufficiently expressive -- and \eflint{}-clingo is sufficiently performant -- to solve real-world problems.
The demonstration is borrowed from a deployment of \eflint{}-clingo in practice.
We briefly describe this usage of \eflint{} and report that all inputs were evaluated sufficiently quickly.

The Products and Data Exchange (PDX) is a product released in October 2025 as part of the Dutch Metropolitan Innovations (DMI) ecosystem for data-driven innovations managed by the Dutch Ministry of Infrastructure.
The PDX serves DMI members as a platform for offering, cataloguing, exchanging, and processing datasets and services.
A novelty of the PDX is its deep integration and automatic enforcement of regulatory documents, in particular the GDPR privacy regulation and the governance rules of the ecosystem.
The governance component of the PDX, developed and maintained by Abykys\footnote{\url{https://abykys.com}}, is built on top of \eflint{}-clingo and enforces the compliance of data-transactions with the regulations formalised in eFLINT.

The `DMI Afsprakenstelsel' defines the rules governing data-transactions and demands compliance with the GDPR.
A precise definition of what is automatically enforced is given at the DMI website\footnote{Here: \url{https://definities.dmi-ecosysteem.nl} (In Dutch).}.
The enforcement of the GDPR focusses on purpose limitation, ensuring that data controller and processor agree on a lawful processing purpose.

\paragraph{Experimental setup}
The specification of the governance rules is continuously extended, refactored, and refined.
We focus on the version that was active on 23 April 2026.
The specification consists of 1,543 lines of code, excluding comments and blank lines.
With Abykys' permission, we have included a sample\footnote{
    Our versions of these test scenarios precisely match the versions made and tested internally by Abykys, except that two strings identifying DMI stakeholders were replaced to protect their privacy.
} of their test scenarios, sampling the spectrum from simple to complex scenarios that Abykys encounters. 
We call these scenarios $s_0$ (the simplest, consisting of zero statements) to $s_3$ (the most complex, consisting of 49 statements).
\Cref{tab:afsprakenstelsel_runtimes} reports on the time required for (stages of) the \eflint{}-clingo interpreter to evaluate each scenario.\footnote{
    As can be seen, these cases all spend the majority of time in the `reasoning' stage. But this belies the time Clingo spends actually applying rules, because this work is interleaved with formatting and writing thousands of lines of answer text as output.
} 
These tests were executed on a machine with an AMD Ryzen AI 7 PRO 350 processor with 8 physical kernels and 16 software threads, 32GB RAM DDR4 memory and an NVMe disc.
The test inputs are available in their entirety at \cite{eflint_afsprakenstelstel_inputs}.
The precise integration of \eflint{}-clingo into the PDX software is not publicly available; we can only report the assurance of the Abykys developers that their version of eFLINT-clingo remains very close to the version openly available at \cite{eflintClingoImpl}.
The same tests were executed with the reference interpreter for a ball-park comparison of total running times.

The experiments in this and the previous subsection served to compare two implementations of eFLINT.
Both implementations deal with the complication that the language is both logical and imperative in nature. 
Specific applications of the language may only use a subset of the language and may thereby permit more efficient execution strategies.
Depending on the features required for an application, choosing an alternative DSL or general-purpose language could therefore lead to better performance than can be realised with the two implementations discussed in this section. 
In future work, we aim to work on eFLINT variants that admit more efficient execution strategies, e.g., by giving a fully procedural interpretation to eFLINT. 
These variants will be more restrictive by giving syntax errors, type errors, or warnings to specifications to which an alternative execution strategy is not applicable.

\bgroup
\setlength\tabcolsep{3pt}
\begin{table}[tb]
    \centering
    \begin{tabular}{lrrrrr|r|r}
        & load & parse & check & trans. & reason & total & v4
        \\
        \hline
        $s_0$ & 35.97 & 34.43 & 13.32 & 7.55 & 318.46 & 409.73 & 15,087
        \\
        $s_1$ & 37.82 & 33.78 & 13.01 & 7.46 & 373.40 & 465.48 & 15,472
        \\
        $s_2$ & 40.28 & 36.47 & 13.65 & 7.58 & 1,055.78 & 1,153.75 & 15,415
        \\
        $s_3$ & 45.87 & 42.07 & 15.07 & 7.62 & 3,603.24 & 3,713.88 & 19,479    
    \end{tabular}
    \caption{Runtime in milliseconds of stages of eFLINT-clingo compared to the eFLINT-v4 total with the \textit{Afsprakenstelsel} as the input specification alongside one of four input scenarios $\{s_0, s_1, s_2, s_3\}$.}
    \label{tab:afsprakenstelsel_runtimes}
\end{table}
\egroup

%% file: clingo_perf1.tex
\begin{figure}[p]
\noindent
\begin{lstlisting}[caption={}]
        Fact x Identified by int Derived from
          (Foreach x: x(x.int - 1) Where 0 < x.int)
\end{lstlisting}%
\begin{tikzpicture}
\begin{axis}[ 
    width=\linewidth,
    line width=0.8,
    mark size=4pt, 
    height=60mm,
    grid=major, 
    xmode=log,           
    ymode=log,           
    log basis x=2,
    log basis y=2,      
    tick label style={font=\small},
    legend style={
        nodes={scale=1.0, transform shape}, at={(0.05,0.95)},
        anchor=north west,
        draw=none,
        fill=none,
        align=left
    },
    legend cell align={left},
    label style={font=\small},
    legend entries={Mercedes star plot, X plot},
    grid style={white},
    xtick=data,
    xlabel={Variable N for scenario \lstinline{+x(}N\lstinline|).|},
    ylabel={run duration (s)},
    y tick label style={
    },
]
\addplot[blue,mark=Mercedes star,dashed,mark options={solid}] coordinates {
(8, 0.387579083442688)
(16, 0.391440749168396)
(32, 0.385046124458313)
(64, 0.39874804019928)
(128, 0.403389930725098)
(256, 0.472855806350708)
(512, 0.748007774353028)
(1024, 1.86181139945984)
(2048, 1.95580208301544)
(4096, 26.6939980983734)
};
\addlegendentry{eFLINT-v4}

\addplot[brown,mark=x,dashed,mark options={solid}] coordinates {
(8, 0.188072681427002)
(16, 0.189316630363464)
(32, 0.193464040756225)
(64, 0.211443305015564)
(128, 0.21407163143158)
(256, 0.244142055511475)
(512, 0.295057892799377)
(1024, 0.389825463294983)
(2048, 0.403687953948975)
(4096, 1.01841914653778)
};
\addlegendentry{eFLINT-clingo} 
\end{axis}
\end{tikzpicture}
\caption{Speed in unfolding long derivation chains.}%
\label{fig:perf_chain}%
\end{figure}

\begin{figure}[p]
\noindent
\begin{lstlisting}[caption={}]
        Fact x Identified by Int Derived from
          (Foreach x1, x2: x((x1 + x2) / 2))
\end{lstlisting}%
\begin{tikzpicture}
\begin{axis}[ 
    width=\linewidth,
    line width=0.8,
    mark size=4pt, 
    height=60mm,
    grid=major, 
    xmode=log,           
    ymode=log,           
    log basis x=2,
    log basis y=2,      
    tick label style={font=\small},
    legend style={
        nodes={scale=1.0, transform shape}, at={(0.05,0.95)},
        anchor=north west,
        draw=none,
        fill=none,
        align=left
    },
    legend cell align={left},
    label style={font=\small},
    legend entries={Mercedes star plot, X plot},
    grid style={white},
    xtick=data,
    xlabel={Variable N for scenario \lstinline{+x(0).+x(N).}},
    ylabel={run duration (s)},
    y tick label style={
    },
]
\addplot[blue,mark=Mercedes star,dashed,mark options={solid}] coordinates {
(0, 0.319804191589356)
(4, 0.42061972618103)
(8, 0.448843002319336)
(16, 0.426221966743469)
(32, 0.422504663467407)
(64, 0.439440846443176)
(128, 0.464979648590088)
(256, 0.566477298736573)
(512, 1.10995721817017)
(1024, 3.36104011535645)
(2048, 3.2697526216507)
(4096, 51.4273458719254)
};
\addlegendentry{eFLINT-v4}

\addplot[brown,mark=x,dashed,mark options={solid}] coordinates {
(0, 0.176027536392212)
(4, 0.181095957756043)
(8, 0.188082337379456)
(16, 0.186144232749939)
(32, 0.188241600990296)
(64, 0.199636101722717)
(128, 0.219603061676026)
(256, 0.274117231369019)
(512, 0.555517196655273)
(1024, 1.77636504173279)
(2048, 1.67821311950684)
(4096, 29.5854017734528)
};
\addlegendentry{eFLINT-clingo} 
\end{axis}
\end{tikzpicture}
\caption{Speed in reasoning with substantial integer arithmetic.}%
\label{fig:perf_arith}%
\end{figure}

\begin{figure}[p]
\noindent
\begin{lstlisting}[caption={}]
Fact x Identified by Int Derived from 
  (Foreach y: y.x1), (Foreach y: y.x2)
  (Foreach y: y.x3), (Foreach x: x(x - 1) Where 0 < x)
Fact y Identified by x1 * x2 * x3 Derived from
  (Foreach x: y(x,x,x)),
  (Foreach x1, x2, x3: y(x1,x2,x3)
                Where (x1 == x2 || x2 != x3)
                   && (Exists y: x2 < y.x1))
\end{lstlisting}%
\begin{tikzpicture}
\begin{axis}[ 
    width=\linewidth,
    line width=0.8,
    mark size=4pt, 
    height=60mm,
    grid=major, 
    ymode=log,           
    log basis y=2,      
    tick label style={font=\small},
    legend style={
        nodes={scale=1.0, transform shape}, at={(0.05,0.95)},
        anchor=north west,
        draw=none,
        fill=none,
        align=left
    },
    legend cell align={left},
    label style={font=\small},
    legend entries={Mercedes star plot, X plot},
    grid style={white},
    xtick=data,
    ylabel={run duration (s)},
    y tick label style={
    },
]
\addplot[blue,mark=Mercedes star,dashed,mark options={solid}] coordinates {
(0, 0.39999794960022)
(1, 0.450055956840516)
(2, 0.448081970214844)
(3, 0.454226016998291)
(4, 0.474435567855835)
(5, 0.487913846969605)
(6, 0.544711589813232)
(7, 0.665695786476136)
(8, 0.905436754226685)
(9, 1.35537552833557)
(10, 2.19583654403687)
(11, 3.57795083522797)
(12, 5.91782748699188)
(13, 9.66319346427918)
(14, 15.4669117927552)
(15, 24.0920568704605)
};
\addlegendentry{eFLINT-v4}

\addplot[brown,mark=x,dashed,mark options={solid}] coordinates {
(0, 0.251827716827392)
(1, 0.261441826820374)
(2, 0.258319497108459)
(3, 0.264661073684693)
(4, 0.280735492706299)
(5, 0.304043292999267)
(6, 0.336050271987915)
(7, 0.378538727760315)
(8, 0.438530564308167)
(9, 0.514181613922119)
(10, 0.596740961074829)
(11, 0.715814232826233)
(12, 0.836026191711426)
(13, 0.975961685180664)
(14, 1.17732620239258)
(15, 1.38846635818481)
};
\addlegendentry{eFLINT-clingo} 
\end{axis}
\end{tikzpicture}
\caption{Speed in reasoning in cases requiring the enumeration, combination, and comparison of many instances.}%
\label{fig:perf_combo}%
\end{figure}

%% file: clingo_perf2.tex
\begin{figure}[p]
\noindent
\begin{lstlisting}[caption={}]
Fact x Identified by Int Derived from 
  (Foreach y: y.x1), (Foreach y: y.x2)
  (Foreach y: y.x3), (Foreach x: x(x - 1) Where 0 < x)
Fact y Identified by x1 * x2 * x3 Derived from
  (Foreach x: y(x,x,x)),
  (Foreach x1, x2, x3: y(x1,x2,x3)
                Where (x1 == x2 || x2 != x3)
                   && (Exists y: x2 < y.x1))
\end{lstlisting}
\begin{tikzpicture}
\begin{axis}[ 
    width=\linewidth,
    line width=0.8,
    mark size=4pt, 
    height=80mm,
    grid=major, 
    xmode=log,           
    ymode=log,           
    log basis x=2,
    log basis y=2,      
    tick label style={font=\small},
    legend style={
        nodes={scale=1.0, transform shape}, at={(0.05,0.95)},
        anchor=north west,
        draw=none,
        fill=none,
        align=left
    },
    legend cell align={left},
    label style={font=\small},
    legend entries={Mercedes star plot, X plot},
    grid style={white},
    xtick=data,
    xlabel={Variable N for scenario of \lstinline{+x(4).} repeated N times},
    ylabel={run duration (s)},
    y tick label style={
    },
]
\addplot[blue,mark=Mercedes star,dashed,mark options={solid}] coordinates {
(0000, 0.387885332107544)
(0001, 0.450289726257325)
(0002, 0.536948919296265)
(0004, 0.673975229263306)
(0008, 0.935039639472962)
(0016, 1.4061849117279)
(0032, 2.44486212730408)
(0064, 4.42739343643189)
(0128, 8.58539271354676)
(0256, 16.8964576721192)
(0512, 32.9207727909088)
(1024, 68.2641124725342)
};
\addlegendentry{eFLINT-v4}

\addplot[brown,mark=x,dashed,mark options={solid}] coordinates {
(0000, 0.277608156204224)
(0001, 0.329833388328553)
(0002, 0.362253308296204)
(0004, 0.432663559913636)
(0008, 0.553332924842835)
(0016, 0.793164730072022)
(0032, 1.27725410461426)
(0064, 2.27849137783051)
(0128, 4.44630753993988)
(0256, 8.84601211547852)
(0512, 17.7607433795929)
(1024, 37.502453327179)
};
\addlegendentry{eFLINT-clingo} 
\end{axis}
\end{tikzpicture}
\caption{Speed in reasoning about the same, moderate reasoning work at each step of scenarios with varying lengths.}
\label{fig:perf_long_time}
\end{figure}

\begin{figure}[p]
\noindent
\begin{lstlisting}[caption={}]
        Fact prime Identified by Int
          Derived from (Foreach int: prime(int)
            Where Not(Exists int1, int2:
              1 Where int1 * int2 == int))
        Event addleq Related to int Creates int
          Syncs with addleq(int - 1) Where 2 < int
\end{lstlisting}
\begin{tikzpicture}
\begin{axis}[ 
    width=\linewidth,
    line width=0.8,
    mark size=4pt, 
    height=96mm,
    grid=major, 
    log basis y=2,      
    tick label style={font=\small},
    legend style={
        nodes={scale=1.0, transform shape}, at={(0.05,0.95)},
        anchor=north west,
        draw=none,
        fill=none,
        align=left
    },
    legend cell align={left},
    label style={font=\small},
    legend entries={Mercedes star plot, X plot},
    grid style={white},
    ylabel={run duration (s)},
    xtick={0,50,100,150,200,250,300},
    xlabel={Variable $N$ for scenario \lstinline{add_leq(N).}}
]
\addplot[blue,mark=Mercedes star,dashed,mark options={solid}] coordinates {
(0000, 0.343814134597778)
(0025, 0.408541679382325)
(0050, 0.542001366615296)
(0075, 0.938855767250061)
(0100, 1.75040078163147)
(0125, 2.96012318134308)
(0150, 4.81333279609681)
(0175, 7.42797052860261)
(0200, 10.9706864356994)
(0225, 15.4166333675385)
(0250, 21.1507836580277)
(0275, 27.9833296537399)
(0300, 36.3604958057404)
};
\addlegendentry{eFLINT-v4}
\addplot[brown,mark=x,dashed,mark options={solid}] coordinates {
(0000, 0.197346091270447)
(0025, 0.373700857162476)
(0050, 0.90993869304657)
(0075, 1.8438937664032)
(0100, 3.20580184459687)
(0125, 4.84339642524719)
(0150, 6.96005260944367)
(0175, 9.527512550354)
(0200, 12.4712575674057)
(0225, 15.9040549993515)
(0250, 19.6240062713623)
(0275, 23.6853358745575)
(0300, 28.3523342609406)
};
\addlegendentry{eFLINT-clingo} 
\end{axis}
\end{tikzpicture}
\caption{Speed in enumerating all prime numbers up to a maximum.}
\label{fig:perf_primes}
\end{figure}

%% file: sections/discussion.tex
\section{Discussion and Future Work}
\label{sec:discussion}
This section reflects on the satisfaction of the requirements of \Cref{tab:requirements1,tab:requirements2}.
References to the requirements are given in \textbf{bold}.
The reflection is a continuation of the discussion around requirements in \Cref{sec:design,sec:implementation} and is based on experience with software systems applying eFLINT.
An overview of the software is given in \Cref{tab:experiments}.
After discussing these experiences, \Cref{sec:future-work} discusses limitations and describes possible avenues of future work.
\input{assets/table_software}

The development of eFLINT went through phases, starting with the analyses of historical cases and simulation, followed by dynamically evolving cases, model checking and system integration.
Applications involving eFLINT have been developed by students, researchers and practitioners at various technological readiness levels (TRLs), with most applications validating concepts `in the lab' ($\leq$TRL3, following the definition of TRL by the European Committee) or performing demonstrations with stakeholders in realistic test environments ($\leq$TRL6).
Especially notable is the ($\sim$TRL8) application of eFLINT as part of the DMI ecosystem mentioned in \Cref{sec:performance-afsprakenstelsel}.
%
%
No applications of eFLINT rely on specific features of a version that make it impossible to migrate to the latest version of the language.

\input{sections/applications}


\subsection{Limitations and Future Work}
\label{sec:future-work}
The various experiments discussed in the previous subsections demonstrate the expressive capability of the language, including the suitability of the \req{datatypes} offered for knowledge representation.
For specific applications, additional primitive types and operators may be needed, such as real or floating point numbers and operators for string manipulation.
Future versions of the language may add additional types and operators, although potentially at the cost that not all implementations offer support.
The semantics of eFLINT can be made more expressive by considering continuous elements, e.g., to represent clocks, and probabilistic reasoning. 

Pragmatic design decisions have been made to ensure the language is sufficiently flexible to be used for various kinds of applications.
However, a flexible solution is not always the most user-friendly solution.
For example, in some applications a strict separation between specification (type declarations) and scenarios (statements) is useful, and the closed world assumption and finite domains are needed for model checking. 
In these applications, the necessary constraints need to be \textit{assumed}.
Conversely, a variant of the language dedicated to an application could \textit{enforce} the constraints syntactically or using static analyses. 
For security reasons, it would also be better to explicitly hide/expose types using a module system (\req{service interface}).

At present, the language is not sufficiently friendly for use by legal experts without software expertise.
Experience with computer science students does suggest that the language is relatively easy to learn for those with programming experience, especially declarative programming experience.
Both language implementations are also lacking a mechanism to report how facts came to be true (or false).
Ideally, such a report contains the sequence of rules and statements that were applied to render the fact true (or false).
This feature is difficult to implement due to reasoning by default (e.g., in case of negation) and the subtle interplay between derivation rules and state transitions.
The information from such a report is essential to achieve \textbf{explainability}.
In \Cref{sec:design} we discussed the use of an execution graph in the reference interpreter to maintain execution history as an essential, yet less fine-grained, contributor to explainability.
Note that in general, however, explainability requires the tailoring of information to specific receivers, tying the requirement together with the \textbf{domain-users} requirement.
That is, legal and software experts appreciate a different perspective on the low-level system information. 

An open question still being investigated is how the \textbf{interpretation} process can best be supported, especially to deal with the amount of existing laws and regulations and the pace with which they are produced.
The current FLINT Rule Editor deploys a pipeline in which natural language processing (NLP) produces FLINT frames from a normative source.
A legal expert then completes, refines and verifies the generated FLINT frames.
A software expert completes and refines the eFLINT code automatically generated from the FLINT frames by using eFLINT's extension mechanism.
In a new project we will experiment with the use of generative AI within this pipeline.

Software expertise is needed, in particular, to refine the eFLINT code in order to formalise the connection between the eFLINT specification and the software environment in which the specification will integrate, e.g., using qualification rules, instrumentation, scripting and/or API adaptation.
Software expertise is also needed to add computational details not easily derived from the natural language sources, such as quantification and constraints on the possible substitutions of variables.
%
%
For consistency, however, it may be preferred for both experts to work on shared artefacts written in the same language.

As possible future work, we envisage a norm specification language which comprises an intermediate language and one or more legal surface-level languages.
The experiences gained during the development of eFLINT, reported in this paper, make it possible to identify a minimal computational core that is complete with respect to the requirements.
One motivation behind such a (re-)design is improving usability for legal and domain experts by offering a more tailored experience and potentially allowing both experts to work on the same specification.
For example, a keyword such as \lstinline-Syncs with- should be hidden as we should not require legal experts to think about transition system semantics.
A rule-based syntax can be used to express the relation between \lstinline+raise-hand+ and \lstinline+place-bid+ without \lstinline-Syncs with- as follows:

{\scriptsize
\begin{verbatim}
raise-hand(bidder) QualifiesAs 
  place-bid(bidder, display.object, price)
  Where price = ...

... is equivalent to ...
\end{verbatim}%
}%
\vspace{-1em}
\begin{lstlisting}[caption={}]
Extend raise-hand Syncs with place-bid
 (bidder=actor,object=display.object, price = ...)
\end{lstlisting}\noindent%
We can go further by enabling such rules to be expressed in a controlled natural language such as a variant of Logical English, developed by Kowalski for, among other things, coding legal rules as logic programs~\cite{logical_english}. 
Another example is RegelSpraak, developed by the Dutch tax administration to formalise parts of the Dutch tax law~\cite{corsius-etal-2021-regelspraak}.
To support a wide range of user-facing languages, additional \textbf{normative concepts} need to be mapped to computational concepts.
For example, eFLINT equates a prohibition to act with the negation of a permission to act.
Rather than forcing this interpretation, an intermediate language could simultaneously support explicit permissions and prohibitions.
Permissions and prohibitions can then conflict, and a priority mechanism is needed to resolve such conflicts.
For example, permissions and prohibitions may be assigned by actors with different levels of authority or may originate from different sources of norms that differ in precedence or specificity. 
FLINT and eFLINT currently lack an explicit mechanism for conflict resolution (\req{norm priorities}).
The eFLINT semantics give an implicit priority to prohibitions over permissions, and to powers over prohibitions, due to how violations are generated and the effects of actions manifest.

A formal operational semantics and a type system would enable a more precise and detailed discussion on the language's design, its properties and possible variants.
At present, neither exists for eFLINT-v4. 
In~\cite{eflint-asp}, a formal semantics has been given to eFLINT-clingo.

%% file: assets/table_software.tex
\begin{table}[bt]
    \caption{An list of software using eFLINT. Source code is available by clicking on the software title in a PDF version of this paper.}
    \label{tab:experiments}
    \centering
    \begin{tabular}{|l|l|l|l|l|}
         \hline 
         \textbf{Title} {\scriptsize (PoC = Proof-of-Concept)} & \textbf{eFLINT-} & \textbf{Refs} \\
         \hline 
         \href{https://gitlab.com/normativesystems/ui/rule-editor}{FLINT Rule Editor} &v1 & - \\
           \href{https://gitlab.com/amdex-fieldlab/use-cases/medicaldata}{Medical Data PoC} & v2 & \cite{binsbergen2021b}\\
            \href{https://github.com/mostafamohajeri/eumas2022-poc}{Normative Advisors PoC} & v2 & \cite{10.1007/978-3-031-20614-6_18}\\
             \href{https://github.com/mostafamohajeri/jurix-notary}{Ex-post Enforcement PoC} & v2 & \cite{liu2024}\\
       \href{https://gitlab.com/enirolf/haskell-implementation}{eFLINT-check PoC} & v2 & \cite{degeus2022} \\
          \href{https://gitlab.com/eflint/demos/case-management-prototype}{Case Management PoC} & v3 & \cite{steketee2024}\\
          \href{https://gitlab.com/jjokarels/prototype-normative-sources-framework}{Normative Sources PoC} & v3 & \cite{jkarels} \\
        \href{https://gitlab.com/amdex-fieldlab/}{AMdEX Fieldlab PoC} & v3 & \cite{amdex-ra}\\
            \href{https://github.com/BraneFramework}{Brane Orchestrator} & v3 & \cite{alsayedkassem_et_al:OASIcs.Commit2Data.2,ALSAYEDKASSEM2025107550, esterhuyse2022} \\
            \href{https://arxiv.org/src/2502.00138v2/anc}{JustAct PoC} & v3 & \cite{justact2024} \\
          \href{https://gitlab.com/eflint/clingo}{eFLINT reasoner} in DMI & clingo & -\\
        \hline 
    \end{tabular}
\end{table}

%% file: sections/applications.tex
%
\subsection{Automating Governance}
\label{sec:discussion-governance}
The eFLINT language has been used in experiments to investigate automating case assessment by governmental organisations based on laws, e.g., for deciding on permits or tax reduction requests.  
An initial experiment with ICTU -- an IT organisation within the Dutch government -- developed the first version of an editor for developing norm specifications and scenarios with FLINT and eFLINT.
The FLINT Rule Editor was used by ICTU domain experts, successfully confirming the usefulness of the languages for developing formalised \textbf{interpretation}s of laws which can subsequently be used for automatic case \textbf{assessment} and \req{micro-simulation}. 
The separation between FLINT and eFLINT in the editor does make it difficult for legal and software experts (\textbf{domain-users}) to work together.
We expect this limitation can be partially mitigated with better tool support delivering a tighter integration.

Motivated by `Rules as Code'~\cite{MohunRoberts2020}, the initial experiment demonstrated how developing machine-readable interpretations of law can promote government \req{transparency} and can aid resolving disputes caused by differences in interpretation.
For these reasons, FLINT establishes \req{source references} between its frames and the original normative sources in a format following the JuriConnect standard~\cite{juriconnect}.
A later project developed a data structure for storing a multitude of interpretations, efficiently interlinking fragments of eFLINT and normative sources~\cite{jkarels}, demonstrating eFLINT's \req{modularity}, \req{extensibility} and \req{versioning}.

In a subsequent experiment, a proof-of-concept (PoC) case management system was developed to assist civil servants in making decisions on cases~\cite{steketee2024}.
The PoC was developed to demonstrate the ability to adapt to changes in laws or policy (\req{dynamic updates}) and to enforce the powers and duties of both civil servant and citizen.
In the PoC, an eFLINT reasoner is \req{instantiated} to keep track of the progress of gathering information and making decisions on a case.
At any time, the system can inform the users of their permitted actions and duties, e.g., to deliver information (citizen) or to make a decision (civil servant).
The user interface of the system is populated with buttons and form elements based on information received by the reasoner about the physical actions and open types of the specification as well as the contents of the knowledge base.
This mechanism makes the system adaptable in the sense that the UI responds to changes in the underlying eFLINT specification. 
The PoC thus benefitted from eFLINT's ability to implicitly specify a \req{service interface} and to handle \req{case input} and produce \req{case output} upon request.
The underlying specification naturally considered an \req{open world} and specified \req{infinite domains} in order to admit, for example, new citizens registering and submitting unpremeditated numerical information. 
%

Two models were experimented with for keeping track of the progress of cases as they develop over time. 
The required runtime information is recorded statefully by the interpreter or within a separate database.
The former favours performance as less information needs to be processed for every request.
The latter favours resilience as reasoner instances can crash without losing runtime state, thus demonstrating \req{persistence}. 
Stateful execution and persistence through an external database are both at odds with adapting to changes in norms, however.
After all, changes in norms might require \req{data migrations} when affecting the fields in type-declarations.
At present we have not demonstrated how this can be achieved with eFLINT.

Another potential application of eFLINT to governance is to conduct \req{macro-simulations} by establishing representative `populations' of hypothetical citizens and analyse sets of cases, such as tax returns, for these populations.
This makes it possible to determine the effects of (e.g., tax) policies on subsets of the wider population while the policy is drafted.
%
Although eFLINT should be able to support this application, we have not yet given a demonstration.

\subsection{Distributed Systems}
%
%
%
In other applications, we have experimented with the automatic enforcement of (privacy) regulations, consortium agreements and data sharing conditions within distributed data exchange systems. 
In a data exchange system, data producers, consumers and service providers collaborate to execute data processing workflows.
We presented a reference architecture for data exchange systems with automatic enforcement of regulations and agreements in~\cite{amdex-ra}.
In the AMdEX Fieldlab PoC, implementing the architecture, compliance is checked and enforced dynamically by a reasoner listening and responding to messages, benefitting from the \req{event-based} nature of eFLINT.
An eFLINT reasoner has also been used by the Brane Orchestrator for planning and distributing compliant data processing tasks~\cite{esterhuyse2022,alsayedkassem_et_al:OASIcs.Commit2Data.2,ALSAYEDKASSEM2025107550}.
The orchestrator can potentially benefit from eFLINT-clingo's capabilities to \req{search} for compliant plans rather than relying on the current trial and error approach.

As inter-domain applications, data exchange systems require mechanisms for cross-domain authentication~\cite{Gommans2009} and consensus.
Consensus is needed on (a) the applied norms, e.g. to establish which interpretation of the GDPR is applied, on (b) policy input, e.g., whether a particular user is affiliated with a member of a consortium, on (c) the transpired events, e.g., whether data was accessed, and (d) on the decisions made by the normative reasoner, e.g., whether authorisation was given.
%
The JustAct PoC was developed to demonstrate a method to communicate about policies, achieve consensus, and for accountable decision making~\cite{esterhuyse2024a, justact2024}.

In inter-domain applications, \req{auditability} is of high importance~\cite{Prinz2022,Shakeri2019}.
To achieve a form of accountability, we can translate eFLINT to a smart contract language in a way that the interpretation is recorded as a smart contract and the assessed scenario is recorded on the ledger. 
We have conducted experiments to confirm the possibility, although a complete, automated translation is not yet available.
Evidence of the possibility is provided by~\cite{symboleo-sc}, presenting a translation from Symboleo~\cite{symboleo2020} to Solidity~\cite{solidity_online,wood2014}.
%

%

%
%

We developed multi-agent systems to experiment with different architectures and setups for automating compliance in distributed systems.
In these experiments, so-called `normative advisors' are agents that reason with norms based on observations communicated by other agents or by the environment~\cite{10.1007/978-3-031-20614-6_18}.
Our approach made it possible to experiment with different models for distributing institutional facts and institutional reasoning and required multiple eFLINT reasoners to co-exist (\req{instantiation}).
One can define a system with a single agent monitoring and assessing the behaviour of all agents in the system.
Alternatively, every agent can have its own normative advisor and use normative reasoning for planning.
Such planning activities by agents benefitted from \req{simulation} of hypothetical scenarios through the \req{forward chaining} applied by the reference interpreter.
The \textbf{search} capability of eFLINT-clingo can be used for \req{backward chaining}, reasoning backwards from a goal state to the current state.

These experiments also included agents for dynamic, \req{ex-post enforcement} that decide whether and how to respond to observed violations~\cite{liu2024}.
These agents may be fully autonomous, or may require a `human in the loop'.
%

In~\cite{binsbergen2021b}, an experiment is reported that demonstrate how eFLINT can be used to gradually specialise a generic law to a concrete, actionable set of policy rules.
In this Medical Data PoC, elements of the GDPR were concretised by the data processing agreement of a medical consortium. 
The resulting specification was operationalised by dynamically generating access control policies for governing the access to data, demonstrating \req{ex-ante enforcement}.
The experiment also demonstrated the ability to formalise normative sources as their own \req{abstraction level}, ensuring that the formalisation of the GDPR fragments can be reused across applications without modification.
The formalisation of the agreement demonstrated the \req{specialisation} of the GDPR provisions to this specific medical context.
The same mechanism was used to concretise \textbf{open-texture terms}, such as `undue delay' in the GDPR, which are deliberately under-specified at that level of abstraction~\cite{governatori2018}.

\subsection{Bounded Model-Checking}
The applications described in the previous subsection are about \emph{concrete} scenarios -- whether actual, historical, or hypothetical -- and may involve the enumeration of large sets of scenarios, e.g., when testing or running simulations.
In other work we applied bounded model-checking to demonstrate the \req{verification} of safety and liveness properties~\cite{degeus2022} via a translation to nuXmv~\cite{cavadaNuXmvSymbolicModel2014}.
In these experiments we reason about \emph{abstract} scenarios, using, for example, linear temporal logic to describe scenarios mixing concrete details and abstract placeholders.
To apply bounded model-checking, eFLINT specifications must have \req{finite domains} such that in every runtime state, only a finite amount of transitions are possible.
%
Bounded model-checking also requires the \req{closed world} assumption.

A goal of this experiment was to show that model-checking can be used in combination with the aforementioned applications, i.e., to verify certain properties before deploying a specification in a case management or data exchange system, without having to maintain multiple versions of the specification (static, \req{ex-ante enforcement}).
To achieve this, we can rely on eFLINT's features to extend a specification by closing open types and \req{reducing domains} to \req{finite domains}.
This way, the same specification can be extended for model-checking and for deployment.

In future work we wish to demonstrate that we can use model checking for static regulatory compliance checking as described in~\Cref{sec:background-compliance} by taking advantage of the separation between physical and institutional actions in eFLINT.
In addition, we aim to investigate the applicability of combinatorial, property-based testing for eFLINT~\cite{goldstein2021}.

%% file: sections/related_work.tex
\section{Related work}
\label{sec:related_work}
In this section we describe work related to automating compliance and compare eFLINT to existing languages.
Several software languages and logics exist to formalise and reason with norms from various sources and within different application areas.
The language and logics differ primarily in their support for fundamental normative concepts, the types of legal documents targeted by the language, the means of execution, and the intended applicatons.

\paragraph{Normative concepts}
Relative to other languages, eFLINT is most similar to languages based on the event calculus~\cite{kowalski1986,sadri1995,russo2002,charalambides2005} such as Symboleo~\cite{symboleo2020} and InstAL~\cite{instAL2016}.
Like eFLINT(-clingo), the Institutional Action Language (InstAL) is a DSL for specifying norms in terms of duties and powers translating to Answer Set Programming (ASP) for execution.
The language has been primarily proposed for modelling institutions and reasoning within multi-agent systems, whereas eFLINT, in addition, has been developed to integrate in software systems and to reason about regulatory compliance of software.

Symboleo and eFLINT are both based on Hohfeld's legal framework~\cite{hohfeld1917fundamental} that emphasises normative relations between actors centred around the normative concepts of power and duty.
%
eFLINT is designed to specify norms from a wide range of legal documents, whereas Symboleo is designed specifically for contracts, embedding notions of contract state in the language.
Instead of relying on built-in state predicates, an eFLINT user could define such predicates themselves.
The original implementation of Symboleo was also based on logic programming (in Prolog, see~\cite{symboleo2020}) and the language has since also been implemented to admit model checking~\cite{symboleo-mc}, access control~\cite{symboleo-ac} and to generate smart contracts~\cite{symboleo-sc}.

Other formal languages for expressing norms are directly based on modal logics such as deontic logic~\cite{herrestad1991} and defeasible logic~\cite{nute2003,governatori2004}.
Deontic logics have been studied extensively for normative reasoning~\cite{Horty2014DeonticSemantics,gabbay2013handbook}. 
Deontic logics admit modalities (operators) for expressing the permission, prohibition and obligation of propositions built from predicates. 
An appealing aspect of deontic logics is that applications of the operators can be nested, making it possible to capture and compare subtle differences in interpretations. 
Moreover, conflicts between permissions and prohibitions can be explicitly expressed.
A downside is the absence of actions, events and temporal reasoning.

For temporal reasoning, an eFLINT specification defines a process model through its action- and event-type declarations.
As such, it is possible to represent changes in facts and norms, and therefore to formalise the normative concept of power. 
A benefit of eFLINT's event-based approach is that checking the compliance of software systems is simplified because systems are inherently event-based.

The lack of temporal reasoning in deontic logics is addressed by Deontic Action Logics (DALs) shifting the perspective from deontic operators on predicates to deontic operators on actions~\cite{deontic_action_logics2025}.
Actions can be composed via parallel or free-choice composition operators.
Permission and prohibition operators are used to determine under which conditions actions are permitted or prohibited. 
LegalRuleML extends RuleML as a markup language for representing and exchanging legal rules based on deontic and defeasible logic.
An interpretation of LegalRuleML rules in terms of a defeasible logic has been given in~\cite{legalruleml_mdl}.
As part of future work, it would be meaningful to demonstrate to which extent eFLINT can be used as a back-end for executing LegalRuleML rules.
In~\cite{10.1007/978-3-031-21541-4_16}, parts of LegalRuleML have been translated to a custom intermediate language to enable reasoning with multiple automated theorem provers with different levels of expressivity.
Temporal logics such as Linear Temporal Logic (LTL) and Computation-Tree Logic (CTL) have been combined with concepts for normative reasoning.
For example, Normative Temporal Logic (NTL) is a temporal logic that replaces the path quantifiers of Computation-Tree Logic (CTL) to express obligations and permissions~\cite{agotnesTemporalLogicNormative2009}.
Temporal Defeasible Logic (TDL) combines defeasibility and temporal logic~\cite{governatoriChangingLegalSystems2010}.
The FIEVeL specification language is used to model institutional policies~\cite{viganoSymbolicModelChecking2007} based on Ordered Many-Sorted First-Order Temporal Logic (OMSFOTL) and uses SPIN~\cite{spin_book} for model checking. 
In~\cite{astefanoaeiSemanticsVerificationNormative2009}, the connection between coordination problems and norm enforcement is formalised using LTL.
The language Revani~\cite{kafalyRevaniRevisingVerifying2016} uses CTL for the specification of privacy norms in the context of sociotechnical systems.

\paragraph{Computational law}
The languages Catala~\cite{catala2021} and RegelSpraak~\cite{corsius-etal-2021-regelspraak} are explicitly targetted for computational law, such as tax laws or social benefits laws.
Computational law is the subset of the law with statutes describing (potentially complex) computations.
%
%
Both languages are explicitly designed to admit co-development of specifications by lawyers and programmers.
The languages are used as part of legal expert systems~\cite{legal_expert_systems}, sharing our motivation of enabling automated governance (see~\Cref{sec:discussion-governance}).

Focussed on computational law, both languages do not directly embed normative concepts such as permission, obligation or power.
A specification essentially describes a collection of functions defined through rules capturing the intricacies of the computations embedded in laws.
The emphasis is on explicating implicit elements and identifying and resolving ambiguities in the law.
In Catala, specific attention is given to the modular and literate development of a specification, allowing earlier definitions to be supplemented by later definitions through an exception mechanism, comparable to eFLINT's type extensions. 
The lessons learnt -- such as on co-development and working with dates~\cite{10.1007/978-3-031-57267-8_16} -- can inspire next versions of eFLINT.

\paragraph{Contracts}
Contracts form a subset of the law that is characterised by a number of properties that are particularly amenable to automation.  
A contract encodes an agreement between two or more parties with an explicitly defined scope and lays out how the normative positions of the involved parties evolve over time, typically in response to events.
As such, a contract can naturally be formalised by a state machine.
This observation forms the basis underneath \textit{smart contracts}.

First introduced by Szabo~\cite{szabo1997} for the exchange of digital assets, smart contracts are now most popular for their usage as scripts executing transactions recorded on a distributed ledger (the blockchain).
Owing to their execution model, smart contracts can guarantee that the exchange of certain digital assets proceeds according to rules encoded in the smart contracts.
Several blockchain platforms exist, often with their own language for developing smart contracts.
But despite their name, smart contract languages are not designed for automating legal compliance and do not support normative concepts directly~\cite{allen2018,luchi_decs}.
Smart contract languages can form an interesting compilation target for normative specification languages, however.
After all, using the ledger, blockchain applications can provide auditability and transparency~\cite{Prinz2022,Shakeri2019,smartaccess}.
As mentioned earlier, Symboleo contracts translate to smart contracts~\cite{symboleo-sc}.
In~\cite{DAzzopardiPS18}, Azzopardi et al. describe an approach for monitoring deontic concepts using smart contracts.

\paragraph{Access and Usage Control}
A significant body of work exists concerning the formalisation, analysis and enforcement of specific kinds of {policies}~\cite{jabal2019} such as access control policies~\cite{xacml3-errata,iannella2018odrl} and network policies~\cite{alshaer2004}.
The Margrave Policy Analyzer tool\footnote{\url{http://www.margrave-tool.org/}} supports multiple formalisms to reason about such policies.
For example, change-impact analysis makes the impact of changes to policies insightful~\cite{fisler2005} and would be useful to normative specification languages.

Many access control models exist~\cite{DBLP:conf/rbac/SandhuM98,DBLP:conf/rbac/Osborn97}.
The most common are Attribute-Based Access Control (ABAC)~\cite{DBLP:journals/computer/HuKF15}, Role-Based Access Control (RBAC)~\cite{DBLP:journals/computer/SandhuCFY96}, and Purpose-Based Access Control (PBAC)~\cite{byun2005purpose}.
The knowledge representation of eFLINT is sufficiently expressive to capture policies based on roles, attributes and purposes.

XACML is a popular ABAC language and the XACML architecture~\cite{xacml3-errata} is widespread as a model for systems integrating ABAC or other forms of access control.
The primary feature of the XACML architecture is its separation of concerns, making it possible for a system administrator to modify policies without other changes to the system. 
An eFLINT reasoner can be used as a policy decision point within the XACML architecture, as demonstrated in~\cite{amdex-ra,binsbergen2021b}.

Whereas access control can enforce permissions and prohibitions, usage control models take a step towards the enforcement of obligations.
That is, policies can specify invariants that need to be maintained for the duration of a data processing action~\cite{park_usage_control}. 
However, many usage conditions in data processing agreements, such as retention periods, cannot be enforced this way, as they survive beyond the processing activity. 
ODRL is popular as a standard for defining usage control policies controlling specific actions on multi-media assets~\cite{iannella2018odrl,DBLP:conf/jurix/Rodriguez-DoncelVG14,DBLP:conf/ruleml/VosKPS19}.
As a representation language, ODRL does not have a fixed semantics\footnote{A working group is specifying semantics for ODRL here:~\url{https://w3c.github.io/odrl/formal-semantics/}.}.
Although ODRL embeds the concept `obligation', it is rather a permission from the normative perspective~\cite{milen_odrl}.


%% file: sections/conclusion.tex
\section{Conclusion}
\label{sec:conclusion}
This paper has reflected on the design and implementations of eFLINT based on requirements extracted from experiments in various application areas. 
%
%
The main novelty of the language is the orientation of the language towards the formalisation of a wide range of normative sources, including many classes of legal documents such as regulations, laws and contracts. 
A specification in the language defines data-types for knowledge representation, process modelling and establishes norms by applying the language's formalisation of the normative concepts of permission, prohibition, obligation and power.
Through specific language constructs, normative sources can be formalised at their own level of abstraction, reused, interconnected and integrated into software systems.
The language's flexibility allows it to be used across application areas such as static and dynamic regulatory compliance checking, auditing, and simulation.
The flexibility is achieved through certain pragmatic design decisions and by embedding software engineering principles such as modularity and extensibility.
The main future developments are to increase the usability of the language, in particular for legal experts, by developing new ways of interacting with the language, including alternative surface-level languages, static analyses, back-ends, and editors.